\documentclass{aa}

\usepackage{graphicx}
\usepackage{txfonts}
\usepackage{float}
\usepackage[colorlinks=true,linkcolor=blue,citecolor=blue,urlcolor=blue,filecolor=blue]{hyperref}
\begin{document}

	\title{Tracing the stellar component of low surface brightness \\
		Milky Way Dwarf Galaxies to their outskirts I: Sextans
	}
	
	\subtitle{}
	
	\author{L. Cicu\'endez\inst{1,2}
		\and
		G. Battaglia\inst{1,2}
		\and
		M. Irwin\inst{3}
		\and
		J.R. Bermejo-Climent\inst{1,2,4}
		\and
		B. McMonigal\inst{5}
		\and
		N.F. Bate\inst{3}
		\and
		G.F. Lewis\inst{5}
		\and
		A.R. Conn\inst{5}
		\and
		T.J.L. de Boer\inst{3}
		\and
		C. Gallart\inst{1,2}	
		\and		
		M. Guglielmo\inst{5}
		\and
		R. Ibata\inst{6}
		\and
		A. McConnachie\inst{7}	
		\and
		E. Tolstoy\inst{8}
		\and
		N. Fernando\inst{5}			
	}
	
	\institute{Instituto de Astrofísica de Canarias (IAC),
		C/Vía Láctea, s/n, 38205, San Cristóbal de la Laguna, Tenerife, Spain
		\\
		\email{lcicuend@iac.es}
		\and
		Departamento de Astrofí­sica, Universidad de La Laguna, 38206, San Cristóbal de la Laguna, Tenerife, Spain
		\and
		Institute of Astronomy, University of Cambridge, Madingley Road, CB3 0HA, Cambridge, United Kingdom
		\and
		Istituto di Astrofisica Spaziale e Fisica Cosmica di Bologna, via Gobetti 101, 40129, Bologna, Italy
		\and
		Sydney Institute for Astronomy, School of Physics A28, The University of Sydney,
		NSW 2006, Sydney, Australia
		\and
		Observatoire astronomique de Strasbourg, Universit\'e de Strasbourg, CNRS, UMR 7550, 11 rue de l'Universit\'e, F-67000 Strasbourg, France
		\and
		NRC Herzberg Institute of Astrophysics, Dominion Astrophysical Observatory, 5071 West Saanich Road, Victoria, BC V9E 2E7, Canada
		\and
		Kapteyn Astronomical Institute, University of Groningen, 9700AV, Groningen, The Netherlands
	}
	
	\date{Received June 27, 2017; accepted September 09, 2017}
	
	\abstract
	{}
	{We present results from deep and very spatially extended CTIO/DECam $g$ and $r$ photometry (reaching out to $\sim$ 2 magnitudes below the oldest main-sequence turn-off and covering $\sim$ 20 deg$^2$) around the Sextans dwarf spheroidal galaxy. We use this data-set to study the structural properties of Sextans overall stellar population and its member stars in different evolutionary phases, as well as to search for possible signs of tidal disturbance from the Milky Way, which would indicate departure from dynamical equilibrium.}
	{We perform the most accurate and quantitative structural analysis to-date of Sextans' stellar components by applying Bayesian MCMC methods to the individual stars' positions. Surface density maps are built by statistically decontaminating the sample through a matched filter analysis of the colour-magnitude diagram, and then analysed for departures from axisymmetry.}
	{Sextans is found to be significantly less spatially extended and more centrally concentrated than early studies suggested. No statistically significant distortions or signs of tidal disturbances were found down to a surface brightness limit of $\sim$ 31.8 mag/arcsec$^{-2}$ in V-band. We identify an overdensity in the central regions that may correspond to previously reported kinematic substructure(s). In agreement with previous findings, old \& metal-poor stars such as Blue Horizonal Branch stars cover a much larger area than stars in other evolutionary phases, and bright Blue Stragglers (BSs) are less spatially extended than faint ones. However, the different spatial distribution of bright and faint BSs appears consistent with the general age/metallicity gradients found in Sextans' stellar component. This is compatible with Sextans BSs having formed by evolution of binaries and not necessarily due to the presence of a central disrupted globular cluster, as suggested in the literature. We provide structural parameters for the various populations analyzed and make publicly available the photometric catalogue of point-sources as well as a catalogue of literature spectroscopic measurements with updated membership probabilities.}
	{}
	
	
	\titlerunning{Tracing the stellar component of low surface brightness MW Dwarf Galaxies out to their outskirts I: Sextans}
	\maketitle

	\section{Introduction}
	
	The faint, passively evolving dwarf galaxies which make up most of the satellite system of the Milky Way (MW), i.e. the dwarf spheroidal galaxies (dSphs), are the subject of a large body of works in the literature, due to their intriguingly large dynamical mass-to-light ratios \citep[see e.g. recent reviews by][and references therein]{battaglia13, walker13} and the possibility they offer to learn about how galaxy evolution proceeded at the low end of the galaxy mass function (see e.g. the review articles by \citeauthor{mateo98} \citeyear{mateo98} and \citeauthor{tolstoy09} \citeyear{tolstoy09},
	and references therein).
	
	Even though they have small physical sizes, the proximity of these galaxies to the MW results in large angular extents, of approximately a few degrees. Hence the study of these systems has particularly flourished since the advent of wide-area imagers and multi-object spectrographs on 4m-10m telescopes, which allow to determine the properties of their	resolved stellar component in great detail out to their very low surface brightness outskirts. This applies in particular to the ``brightest'' of the MW dSphs\footnote{For convenience, we will also refer to them as to the ``classical dSphs'', i.e. those known before SDSS.}, those with luminosities $-15 \lesssim M_V \lesssim -8$, for which one can gather wide-area photometric data-sets with large number statistics reaching well below the oldest main-sequence turn-off (oMSTO) and spectroscopic data-sets providing velocities and metallicities of several hundreds stars per galaxy.
	
	Thanks to these types of data-sets, the stellar population of several classical MW dSphs is known to exhibit spatial variations both in its age and metallicity properties \citep[][to mention only a few]{harbeck01, tolstoy04, battaglia06, faria07, battaglia11, deboer12, battaglia12}, which point to the central regions having experienced a more prolonged star formation and chemical enrichment history than the outer parts.
	
	Substructures have been found in the density maps and kinematics of some of these systems \citep[e.g. ][]{irwin95, bellazzini02, palma03, wilkinson04, battaglia06, walker06}, possibly due to the disruption of stellar clusters or even smaller accreted dwarf galaxies, where	this informs us on stellar cluster formation in small systems and accretion/merging between low mass haloes (\citeauthor{pace14} \citeyear{pace14}, see also \citeauthor{amorisco14} \citeyear{amorisco14} for the M31 satellite And~II). 
	
	Chemo-dynamical stellar components have been detected \citep{tolstoy04, battaglia06, battaglia11}, and their simultaneous mass modeling has been proposed as a way to relieve known degeneracies in the Jeans analysis of spherical systems \citep{battaglia08}	and place further constraints on their dark matter density profile \citep[e.g.][]{walker11, amorisco2012fnx}. The large number statistics and accurate radial velocities provided by available spectroscopic data-sets of individual stars in MW dSphs has also allowed the use of	sophisticated mass modeling techniques \citep[e.g.][]{breddels13, zhu16}. The dynamical modeling of MW dSphs relies on the assumption that these systems are in dynamical equilibrium at all radii. However, elongations and/or distortions in the outer parts of the stellar body have been found in a handful of MW dSphs, suggesting that the present-day properties of the stellar component of these specific galaxies might be affected by tidal forces exerted by the Milky Way \citep[e.g.][]{bellazzini02, battaglia12, mcmonigal14, roderick15, roderick16boo}. Depending on the degree of tidal stripping, this might have implications
	for estimates of the dark matter halo properties \citep[e.g.][]{munoz08}. Strong tidal disturbances have also been shown to attenuate initial differences in the spatial and kinematic properties of stellar populations, offering a possible explanation as to why in some dSphs the presence of chemo-dynamical stellar components is not as evident as in other ones,
	such as in Carina \citep[e.g.][]{sales10}. 
	
	In this work, we study in detail the properties of the stellar component of the MW satellite Sextans, in search for the possible presence of signs of tidal disturbance, to quantify the structural properties of the overall stellar population and	to constrain spatial variations in the properties of the stellar population mix as traced by stars found in different evolutionary phases. 
	
	Discovered in the UKST sky survey by \cite{irwin90}, Sextans is a particularly intriguing object because it was found to have a very low central surface brightness ($\mu_V=27.1$ mag/arcsec$^{-2}$) and a much larger extent than other similarly luminous MW classical dSphs (a King tidal radius of 160 arcmin, corresponding to $r_t=4.0$ kpc, assuming an heliocentric distance of $D_{\odot}=86$ kpc by \cite{lee09}; to be compared for example to King tidal radii of 1-2 kpc for Ursa Minor and Sculptor). These characteristics make Sextans a candidate for having experienced	strong tidal disturbance from the MW, but at the same time hard to study in detail, due to the difficulties of mapping its properties over a large portion of its stellar body and of separating stars belonging to Sextans from the very numerous contaminants (e.g. foreground Milky Way stars).
	
	Its structural parameters were carefully derived for the first time in \cite{irwin95} (hereafter IH95) from photographic plates, but it has been only very recently that this type of analysis has been carried out on deep photometry fully mapping the galaxy out to its outskirts \citep[][hereafter R16]{roderick16} or along portions of its major and minor axis \citep[][]{okamoto17}. The surface density map by R16 exhibit significant distortions in the outer parts, although the main conclusion of the authors was that Sextans is not undergoing strong tidal disruption. R16 revised Sextans nominal King tidal radius down to a much smaller value of 83 arcmin, while \cite{okamoto17} obtained a value of 120 arcmin, although the limited spatial coverage of their data-set makes the determination uncertain. 
	
	The properties of Sextans' stellar population mix are known to vary spatially. By analyzing the colour-magnitude diagram of the central $\sim$ 33 $\times$ 34 arcmin$^2$ of Sextans, \cite{bellazzini01} found evidence for the presence of at least two components in the old stellar population of this galaxy, with the main one having [Fe/H] $\sim -1.8$ and a minor component around [Fe/H] $\sim -2.3$, and hints that the blue horizontal branch (BHB) stars are less spatially concentrated than the stars in the other evolutionary phases analyzed. Later, \cite{lee03}, R16 and \cite{okamoto17} confirmed the larger extent of BHB stars, although they do not	provide a quantification of the different spatial distribution of the stellar populations in terms of structural parameters.
	
	Also the properties of Sextans BSs exhibit spatial variations, with the bright (more massive) blue stragglers being more centrally concentrated than the fainter (less massive) ones \citep{lee03}. \cite{kleyna04} found a cold kinematic substructure close to the center of the Sextans dSph	(see also \citeauthor{battaglia11} \citeyear{battaglia11} and \citeauthor{walker06} \citeyear{walker06}, the latter for a substructure close to Sextans core radius) and suggested that the central cold substructure, the spatial distribution of bright and faint BSs and the sharp central rise in the light distribution of Sextans (e.g. IH95), could be explained by the dissolution of a stellar cluster in Sextans center.
	
	In this article we present an extensive wide-field study of Sextans' stellar population using deep $g-$ and $r-$ band photometry (reaching down to $\sim$ 2 magnitudes below the oldest MSTO) from a mosaic of CTIO/DECam pointings over $\sim$ 20 deg$^2$ along the line-of-sight to the Sextans dSph. This reaches well beyond the nominal IH95 King tidal radius. We perform a detailed and quantitative analysis of the structural properties of this galaxy, using statistically sophisticated tools. The article is organized as follows. Section \ref{sec:observations} presents the details of our observations, the data reduction process and the resulting photometric catalogue, which we make publicly available. In Sect.\,\ref{sec:structural} we extract the structural parameters of the overall stellar population of the galaxy, evaluate the goodness-of-fit of different functional forms for the surface density profile and measure the integrated magnitude and central surface brightness. In Sect.~\ref{sec:maps} we derive Sextans decontaminated surface density map and its deviations from axi-symmetry. The analyses carried out over the whole population of Sextans are applied to the stars in different evolutionary phases in Sect.\,\ref{sec:evol_phase}. In Sect.\,\ref{sec:spectroscopy}, we use our results to update membership probabilities of stars from the \citet{walker09} and
	\cite{battaglia11} spectroscopic samples and make the resulting catalogue publicly available. Finally, Sect.\,\ref{sec:conclusions} is dedicated to the summary and conclusions of this work. 
	
	\section{Observations and Data Reduction}
	\label{sec:observations}
	
	The observations took place between March 18 and 23 2015 in visitor mode with the instrument DECam on the 4m Blanco telescope at the CTIO (PI: B. McMonigal). DECam is a wide-field CCD imager containing 62 2048$\times$4096 pixel CCDs, producing images with a field of view of 2.2 degrees at 0.263 arcsecond/pixel resolution.
	
\begin{figure}
		\centering
		\includegraphics[clip,width=\hsize]{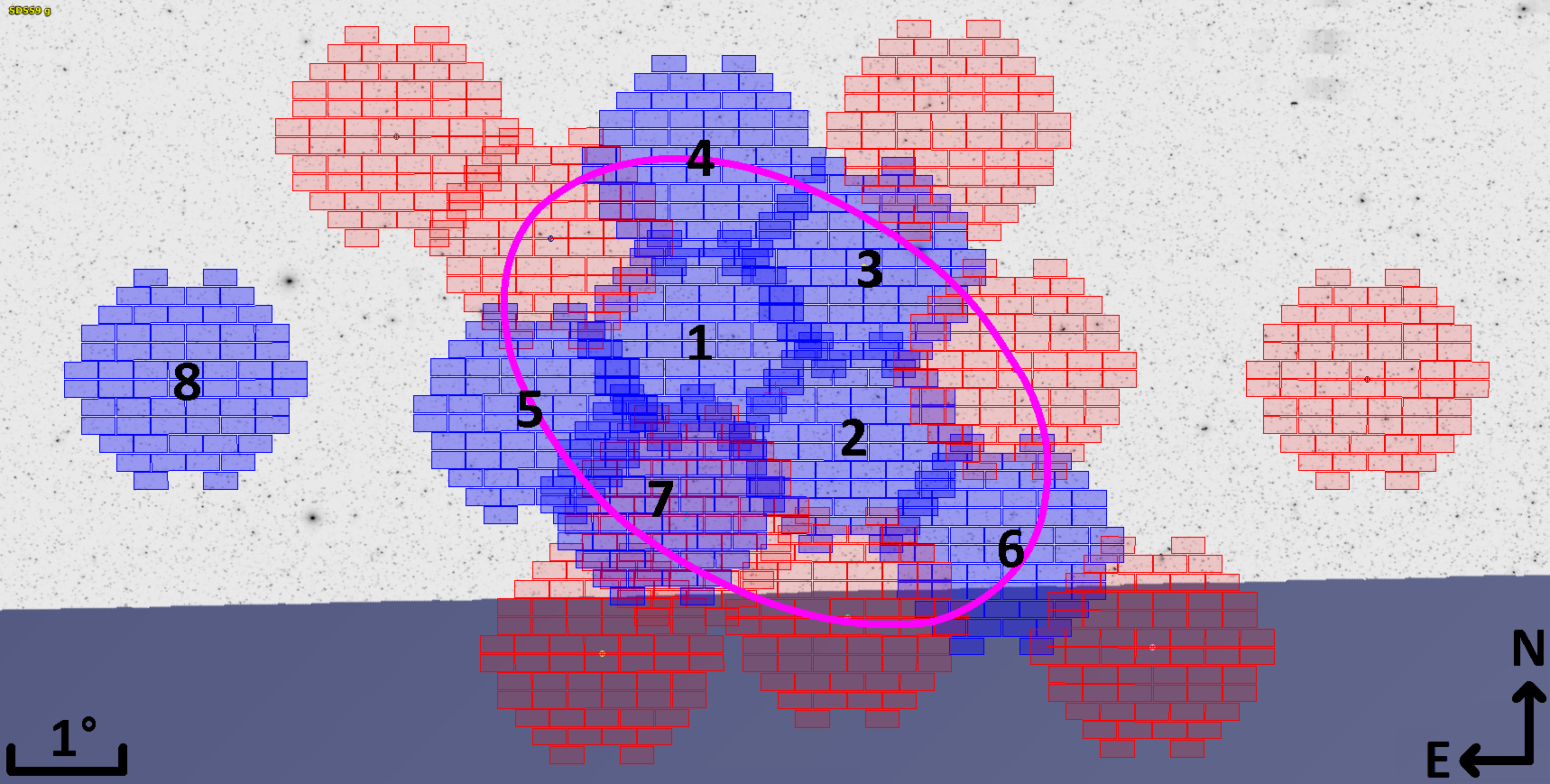}
		\caption{Location of the observed (numbered) and planned (not numbered) DECam pointings around the Sextans dSph, overlaid to SDSS DR9 $g$ band imaging, which covers almost entirely the surveyed area (the blue band is the region lacking SDSS DR9 coverage). Pink ellipse: Contour of the previously estimated King surface density profile of Sextans at its tidal radius. Pointing 7 is from archive DECam observations. The DECam FoV used to prepare this image in Aladin Sky Atlas (\citeauthor{aladin} \citeyear{aladin}) was designed by L. Cicuéndez.}
		\label{locations}
	\end{figure}

	The initial plan for the observations consisted of a mosaic composed of 14 pointings centered on the Sextans dSph and probing out to twice its nominal King tidal radius (adopting the structural parameters derived by IH95), plus 2 displaced ones in order to measure the contamination density (Fig.\,\ref{locations}). However, because of bad weather conditions, only six pointings from the mosaic could be observed (\#1$-$\#6 in Fig.\,\ref{locations}), plus one displaced pointing (\#8). In order to enlarge the area covered along the projected minor axis, we added one pointing (\#7) from the public data at the NOAO Science Archive, forming part of the proposal 2013A-0611 (PI: A.D. Mackey).
	The complete map covers approximately 20 deg$^2$, reaching out to slightly beyond the IH95 estimate of Sextans' nominal King tidal radius.
	
	All the pointings but \#7 were observed with multiple exposures of 300s $g-$ and 500s in $r-$ band, dithering with spatial offsets between the individual exposures chosen to fill the gaps between CCDs. On the other hand, pointing 7 had been observed with 3x300s exposures in $g-$ and $i-$ bands, without dithering. The log of the observed fields is shown in Table \ref{log}.
	
	\begin{table*}
		\caption[.]{Log of the observed fields combined into the analysed map.}             
		\label{log}      
		\centering                          
		\begin{tabular}{c c c c c c c c c}        
			\hline\hline                 
			\noalign{\smallskip}
			Pointing & $\alpha_{2000}$ & $\delta_{2000}$ & UT date & Filter & Average sec z & Average seeing\\    
			& (hh:mm:ss) & (dd:mm:ss) & (dd:mm:yyyy) &  &  & ($''$)\\
			\noalign{\smallskip}
			\hline                        
			\noalign{\smallskip}
			1 & 10:15:44.17 & -01:09:42.39 & 19/03/2015 & g & 1.53 & 1.22 \\      
			& & & & r & 1.36 & 0.98\\
			2 & 10:10:21.84 & -02:04:03.61 & 19/03/2015 & g & 1.22 & 0.98 \\
			& & & & r & 1.16 & 0.97\\
			3 & 10:09:54.72 & -00:30:58.14 & 19/03/2015 & g & 1.19 & 1.19 \\
			& & & & r & 1.35 & 1.38\\
			4 & 10:15:34.80 & +00:23:24.00 & 22/03/2015 & g & 1.17 & 1.25 \\
			& & & & r & 1.16 & 1.17\\
			5 & 10:21:33.60 & -01:48:36.80 & 23/03/2015 & g & 1.49 & 1.01 \\
			& & & & r & 1.33 & 0.92\\
			6 & 10:04:59.52 & -02:58:24.60 & 23/03/2015 & g & 1.17 & 0.98\\
			& & & & r & 1.14 & 0.88\\		
			7 & 10:17:03.30 & -02:31:53.00 & 16/02/2013 & g & 1.22$^1$ & 1.43 \\
			& & & & i & 1.27$^1$ & 1.33\\
			8 & 10:34:00.00 & -01:30:00.00 & 19/03/2015 & g & 1.15 & 1.08 \\
			& & & & r & 1.14 & 0.94\\
			\noalign{\smallskip}
			\hline            
		\end{tabular}
	\end{table*}
	\begin{table*}
		\begin{tabular}{l}
			$^1$Via \url{www.eso.org/sci/observing/tools/calendar/airmass}
		\end{tabular}
	\end{table*}
	
	The data were processed by the Cambridge Astronomical Survey Unit (CASU), using the pipeline described in \cite{irwin01} in an optimized form for working with DECam data. This pipeline performs standard reductions (debiasing, flat-fielding, astrometry and internal photometric calibration between pointings), classifies the detected objects morphologically and generates their corresponding photometric catalog. The morphological classification was done by allowing the point spread function (PSF) to be determined independently per CCD and filter.
	
	Even though we do not perform artificial star tests, we are confident that our data do not suffer
	from crowding: data-sets suffering from crowding will show a shallower limiting magnitude in the densest regions, as well as an increase of sources classified as extended/blends. We have verified that in the innermost regions the spatial distributions of extended and noise-like objects does not show any such feature, as well as that the limiting magnitude of the data in the central region compares well to the rest of the data-set.
	
	Hereafter, unless otherwise stated, we will only refer to objects classified as point-like in both photometric bands to exclude background galaxies, blends or noise detections. Given the comparable depth of the $g-$ and $r-$ photometry, information from both bands should provide a more reliable morphological classification. For each point-like object the pipeline assigned a morphological classification flag per photometric band whose possible values are -1, -2 or -3, which correspond to an identification as an almost certain star, a probable star or a star/compact galaxy respectively.
	
\begin{figure}
		\centering
		\includegraphics[trim={1.75cm 0cm 5.5cm 0.25cm},clip,width=\hsize]{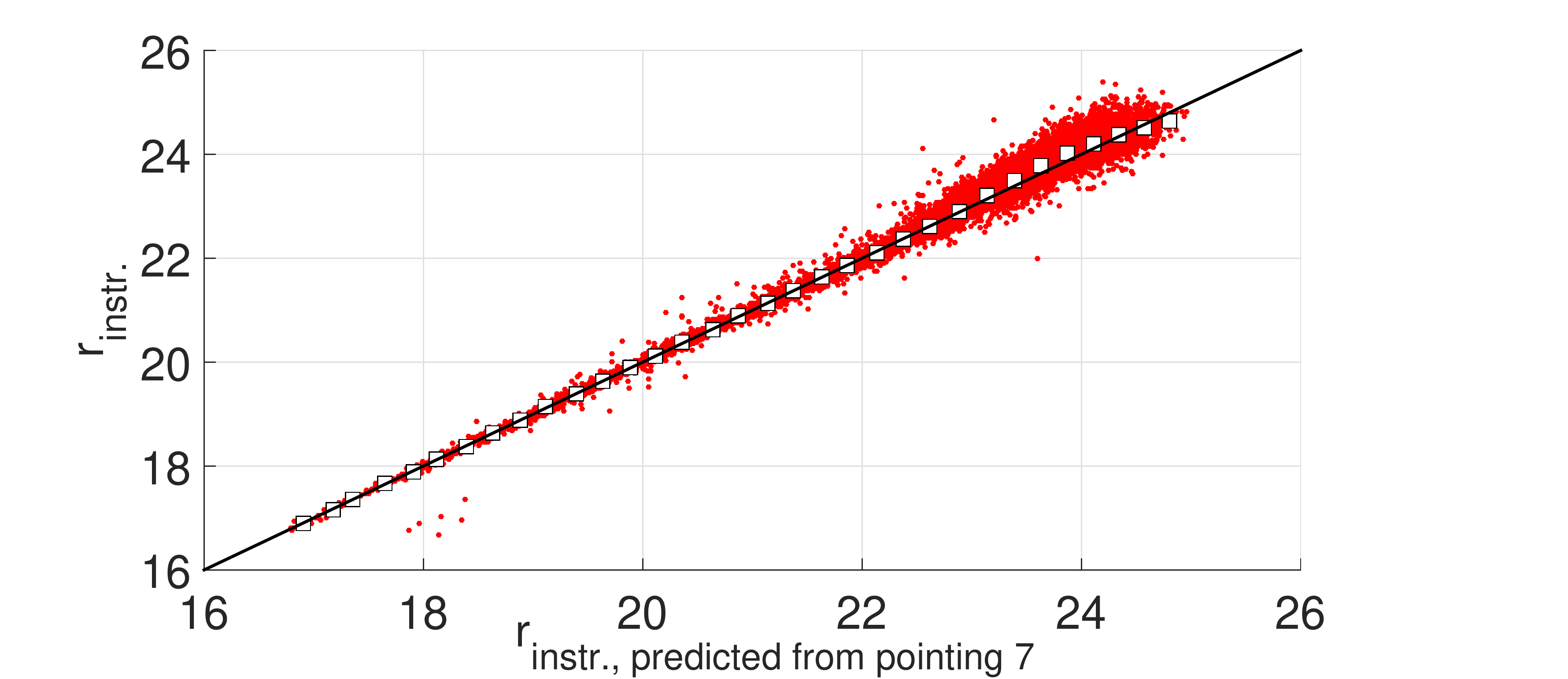}
		\includegraphics[trim={1.75cm 0cm 5.5cm 0.25cm},clip,width=\hsize]{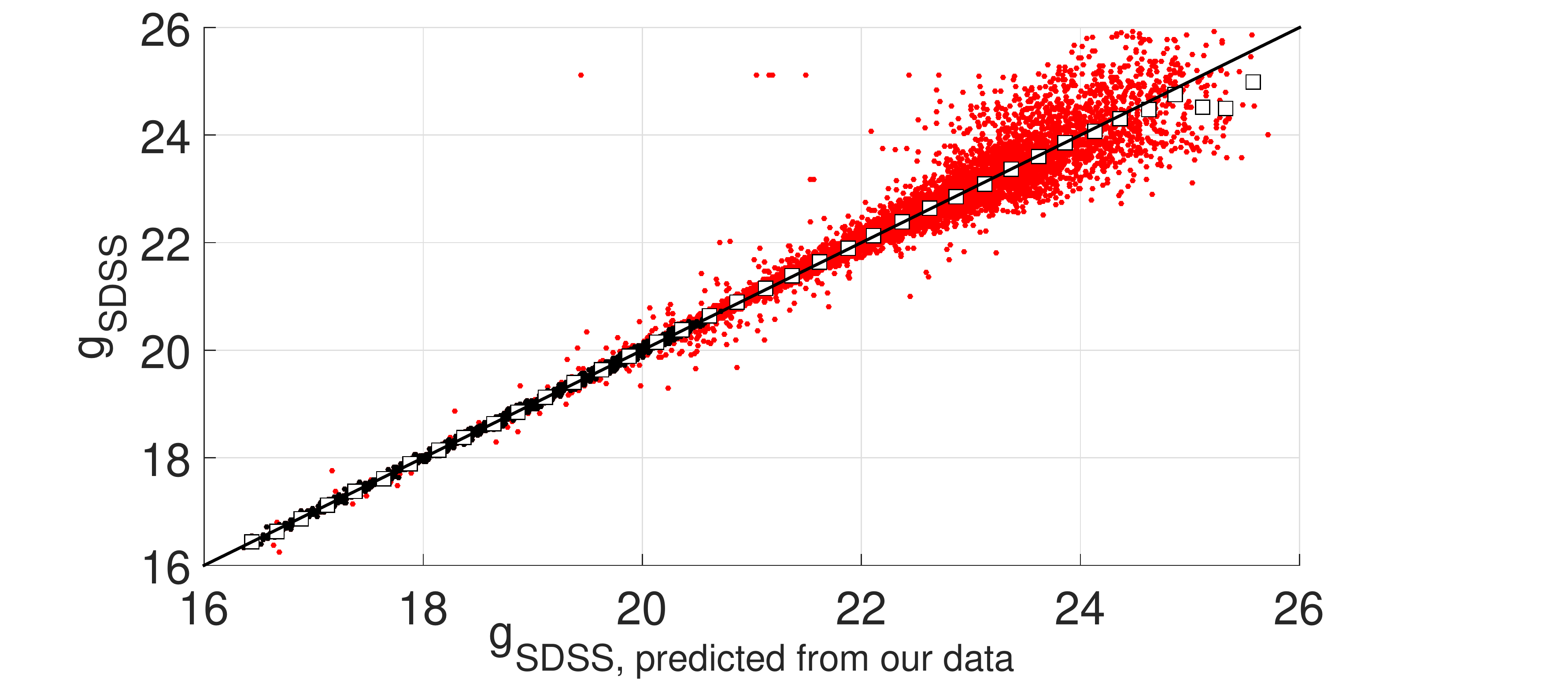}
		\caption{Top: Linearity check of the photometric conversion to $r_{instr.}$ of pointing 7 through its overlapping regions with the rest of the mosaic. Bottom: Linearity check of the photometric calibration with SDSS, in this case for pointings from the mosaic in g band. The stars used for the linear fitting are shown as black dots, and are the majority of those having $g_{\rm SDSS}\lesssim20.5$, while the stars not used are shown as red dots (outliers or those with errors larger than 0.03 mag). In both panels the white squares give the median values in 0.25 mag-wide bins and the black line shows the one-to-one relation.}
		\label{calibration}
	\end{figure}

	The photometry of the mosaic was first calibrated internally (in instrumental magnitudes) using the overlapping regions between pointings; obviously this step could not be applied to the displaced pointing. As pointing 7 was originally observed in $g$ and $i$ bands, for its initial internal calibration with the rest of the mosaic we inferred the $r_{instr.}$ band magnitudes by applying the linear equation $r_{instr.}-g_{instr.} = c_r\,(g_{instr.}-i_{instr.})+ZP_r$ (Fig.\,\ref{calibration}) to the objects overlapping between our mosaic and the one from the original proposal of pointing 7; in this case, we find a colour term $c_r=-0.7101$, with a scatter of $\sigma_r=0.0229$ mag around the relation. The agreement between the measured $r_{instr.}$ band magnitudes and those predicted by the relation is very good, with only a slight deviation for the faintest sources (see top panel Fig.\,\ref{calibration}). 
	
	The photometric calibration was performed by cross-correlating our DECam catalogue with the SDSS DR12 point-like catalogue (\citeauthor{sdssdr12} \citeyear{sdssdr12}) in $g_{SDSS}$ and $r_{SDSS}$ bands, for the mosaic area at once and for the displaced pointing independently. A linear fitting was sufficient for the purpose (Fig.\,\ref{calibration}): 
	{\setlength{\mathindent}{-0.18cm}
		\begin{equation}
		\begin{array}{l}
		g_{instr.}-g_{SDSS} = c_g\,(g_{instr.}-r_{instr.})+ZP_g\\
		r_{instr.}-r_{SDSS} = c_r\,(g_{instr.}-r_{instr.})+ZP_r
		\end{array} 
		\end{equation}}
	
	In order to reject outliers, we applied the bisquare weights method inside a loop to recursively discard objects beyond $3\sigma$ from the linear fitting in each iteration, until convergence. The standard deviation was calculated from the median absolute deviation (MAD) through the expression: $\sigma=\frac{MAD}{\phi^{-1}(3/4)}\simeq 1.4826\,\cdot\,$MAD, where $\phi^{-1}$ is the quantile function for the standard normal distribution. The resulting colour terms are $c_g=-0.0974$ and $c_r=-0.102$, with a scatter of $\sigma_g=0.030$ mag and $\sigma_r=0.024$ mag. The bottom panel of Fig.\,\ref{calibration} shows that the transformation holds well down to the faintest magnitudes. The positions of the stars retained by the linear fitting after rejecting all the outliers through this cleaning method agree with the SDSS astrometry within $\sim$ 0.3$''$.
	
	The depth of the data-set is M$_{g,\,AB}\simeq25.1$ and M$_{r,\,AB}\simeq24.9$ at S/N $\simeq 5$, M$_{g,\,AB}\simeq25.8$ and M$_{r,\,AB}\simeq25.4$ at S/N $\simeq 3$.
	
	We correct the photometrically calibrated $g-$ and $r-$ magnitudes for Galactic extinction using the \cite{schlegel98} dust map together with its 14\% reddening recalibration and extinction coefficients derived by \cite{schafly11}. Due to variations of the extinction across the large observed area, we correct it star by star, limited by the spatial resolution of the dust map. For our observed fields the median values of the extinction were $A_g=0.144$ mag and $A_r=0.100$ mag, the maximum were $A_g=0.247$ mag and $A_r=0.171$ mag and the minimum $A_g=0.080$ mag and $A_r=0.056$ mag. Hereafter, we only show the extinction and reddening corrected photometry. The catalogue of point-like sources is given in Table \ref{photometric_catalogue}.
	
	\begin{table*}
		\caption[.]{A sample of the Sextans point-like catalogue in SDSS photometric system. The classification flags in each band are described in Sect.\,\ref{sec:observations}. The reddening E(B-V) was derived from \cite{schlegel98} dust map. The full photometric catalogue is available online together with the electronic version of this article.}
		\label{photometric_catalogue}
		\centering
		\begin{tabular}{r c c c c c c c}
			\hline\hline
			\noalign{\smallskip}
			ID & $\alpha_{2000}$ & $\delta_{2000}$ & g & g class & r & r class & E(B-V)\\    
			& (hh:mm:ss) & (dd:mm:ss) & (mag) &  & (mag) &  & (mag)\\
			\noalign{\smallskip}
			\hline
			\noalign{\smallskip}
			1  &  10:13:29.2581 & -01:11:4.808 & 24.736$\pm$0.199 & -1 & 24.870$\pm$0.160 & -1 & 0.03966\\
			2  &  10:14:02.1465 & -01:11:5.040 & 21.318$\pm$0.047 & -1 & 19.874$\pm$0.025 & -1 & 0.04122\\		
			3  &  10:13:37.9707 & -01:11:5.391 & 24.285$\pm$0.141 & -1 & 24.400$\pm$0.109 & -3 & 0.03982\\		
			4  &  10:13:57.1101 & -01:11:7.043 & 24.411$\pm$0.145 & -2 & 24.325$\pm$0.097 & -3 & 0.04205\\		
			5  &  10:13:55.6720 & -01:11:7.195 & 24.427$\pm$0.145 & -2 & 24.124$\pm$0.081 & -1 & 0.04205\\
			6  &  10:14:32.6120 & -01:11:8.072 & 24.693$\pm$0.171 & -1 & 23.898$\pm$0.067 & -3 & 0.04309\\
			7  &  10:14:29.1699 & -01:11:8.671 & 25.059$\pm$0.241 & -1 & 25.054$\pm$0.176 & -1 & 0.04361\\
			8  &  10:14:34.0414 & -01:11:8.968 & 22.127$\pm$0.050 & -1 & 21.948$\pm$0.028 & -1 & 0.04309\\
			9  &  10:13:37.0119 & -01:11:8.972 & 24.576$\pm$0.167 & -3 & 24.003$\pm$0.077 & -1 & 0.03982\\
			10  &  10:14:17.7718 & -01:11:9.697 & 23.896$\pm$0.091 & -1 & 22.242$\pm$0.029 & -1 & 0.04281\\					
			... & ... & ... & ... & ... & ... & ... & ...\\
			\noalign{\smallskip}
			\hline   
		\end{tabular}
	\end{table*}
	
\begin{figure*}
		\centering
		\includegraphics[trim={0.5cm 0.5cm 16cm 0.25cm},clip,width=\hsize]{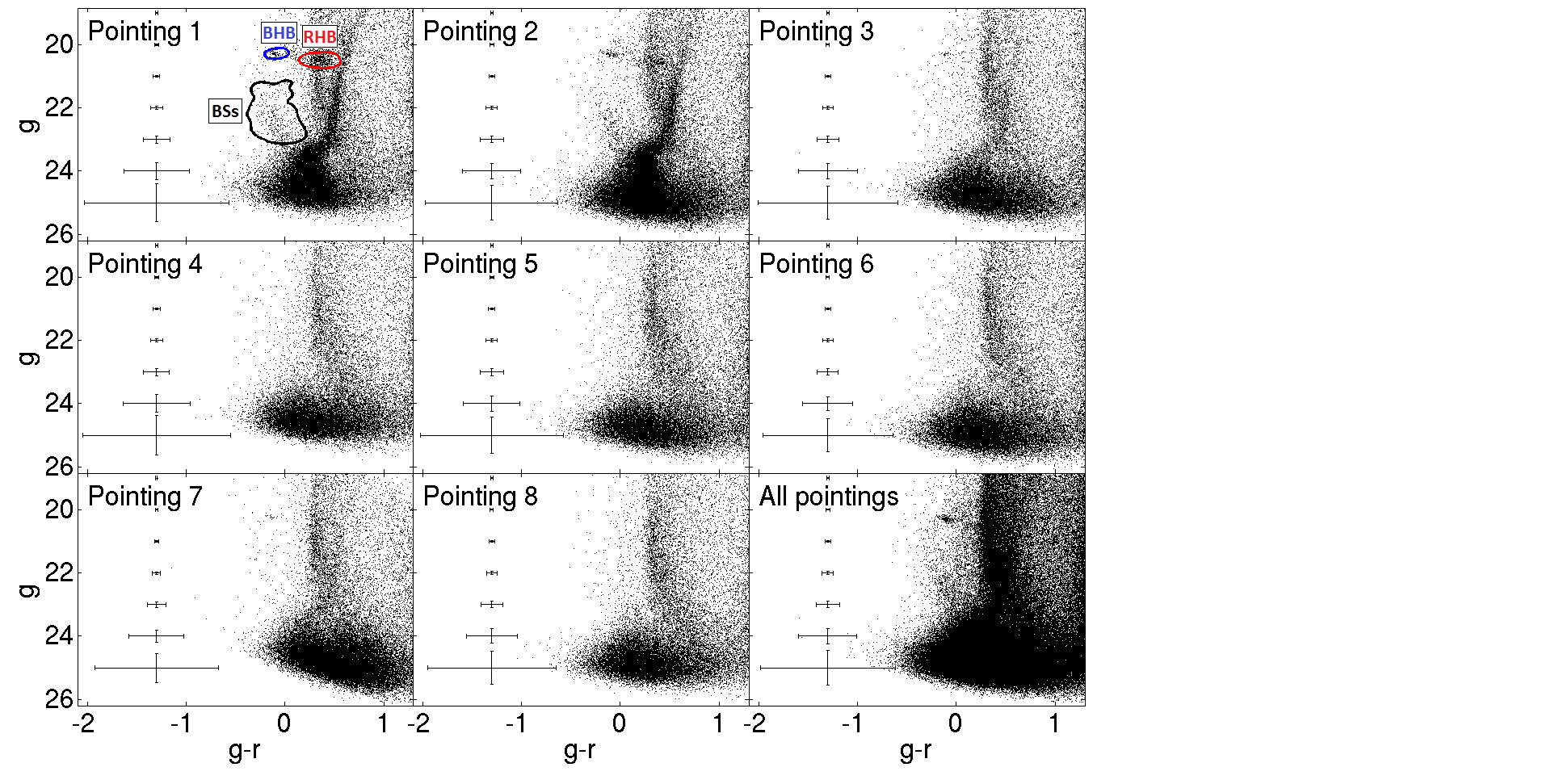} 
		\caption{CMDs of the different DECam pointings, together with the representative 3$\sigma$ errors in mag and colour (errorbars on the left). MS, MSTO, BSs candidates, RGB, RHB and BHB features from Sextans stellar population are clearly visible in pointings 1 and 2, while in 3, 7 are less evident and in the rest they are indistinguishable from the contamination (remember that \#8 is assumed to contain just contamination). In pointing \#1 we also show the selection windows used to isolate RHB, BHB and BS stars later on. Pointing \#4 is the shallowest pointing and \#6 the deepest; \#8 is displaced from the main mosaic area for foreground/background determination.}
		\label{cmds}
	\end{figure*}

	The CMDs of the individual pointings are shown in Fig.\,\ref{cmds}. The features of Sextans stellar population (main sequence, main sequence turn-off, sub- and red giant branch, blue stragglers, red and blue horizontal branch) are clearly recognizable in field 1 and 2, becoming less evident in 3\&7, while the rest of the fields appear to contain only Milky Way stars and unresolved background galaxies. 
	
	Among the aims of our work is to explore the 2D structural properties of the Sextans stellar population, both by determining Sextans structural parameters and surface density profile, and by investigating the possible presence of signs of tidal disturbance, such as e.g. low surface density tidal debris or structural irregularities.
	
	However, on instruments with such a large field of view as DECam it is common that some regions are out of focus, in particular in the outer parts. The locations of these regions depend on the optical aberrations of the telescope, but also on factors that change from pointing to pointing, like the focus carried out at the beginning of each night or the inclination of the telescope, affecting the tilt of the focal plane or the tension on the mirrors. These distortions in the focal plane can lead to a morphological misclassification (see e.g. \citeauthor{mcmonigal14} \citeyear{mcmonigal14}) whose importance depends on such variables as the depth, the seeing, etc. which once again change from pointing to pointing. As a result, in the most out-of-focus regions extended objects tend to be detected as point-like ones due to our treatment of the point-spread-function across the field-of-view, resulting in artificial overdensities of point-like sources. In our case, these features become noticeable when including objects fainter than $(g,r)=(23.0,23.0)$ mag, which is approximately the region where the locus of unresolved galaxies starts appearing on the CMD. This is clearly something that stands in the way of detecting low surface brightness features in surface density maps, hence in the following, unless specified otherwise, we perform our analysis on objects brighter than $(g,r)=(23.0,23.0)$ mag. In Appendix A we discuss the various attempts we made to overcome this issue. The chosen magnitude cut also ensures that all the pointings are at the $\sim$ 100\% completeness level; we have verified this statement by examining the ratio of the luminosity function of Milky Way dwarf stars at $1.1<g-r<1.6$ with the predictions from the Besan\c con model\footnote{\url{http://model.obs-besancon.fr}} (\citeauthor{besancon03} \citeyear{besancon03}): in the magnitude range considered, the ratio is constant around unity and then shows a clear, sharp drop-off at fainter magnitudes, around $\sim$24 mag, when the completeness starts decaying.

	\section{Structural parameters}
	\label{sec:structural}
	
	In order to derive the structural parameters of the Sextans dSph whilst avoiding the loss of information due to spatial binning of the data, we used the method defined in the appendix of \cite{richardson11}. This method works by evaluating likelihoods at each star’s location given the expression of the galaxy surface density profile, which are also a function of the galaxy structural parameters (e.g. centre, ellipticity, position angle, etc.), and contamination density.
	
	The likelihood of observing the $N$ data points at positions $\textbf{r}_\textbf{i}$, with $i=1,2,...,N$, for a given surface density profile $f(\textbf{r})$ (dSph + contamination) is then:
	\begin{equation}
	\ln L = \int_S \! f(\textbf{r}) \, \mathrm{d}S + \sum\limits_{i=1}^N ln \, f(\textbf{r}_\textbf{i})\,,
	\label{eq:ln_L}
	\end{equation}
	with the integral evaluated over the observed area $S$; $\textbf{r}_\textbf{i}$ is the star's elliptical radius (the semi-major axis of the ellipse passing at the location of the star $i$) projected onto the tangent plane of the sky at the galaxy centre (in standard coordinates:  $\xi,\eta$):
	\begin{equation}
	\textbf{r}_\textbf{i}^2 = (\Delta \xi_i \sin \theta+\Delta \eta_i \cos \theta)^2+(-\Delta \xi_i \cos \theta+\Delta \eta_i \sin \theta)^2/(1-\epsilon)^2\,,
	\end{equation}
	with $(\Delta \xi_i,\Delta \eta_i)$ being the star' displacement with respect to the galaxy centre, $\theta$ the position angle (measured North through East) of the ellipse and $\epsilon$ its ellipticity\footnote{It is worth mentioning that one term is missing on the right-hand side of Eq.\,\ref{eq:ln_L} from \cite{richardson11}, i.e. $+\sum_{i=1}^{N}\ln dS$, evaluated over the stars' locations. This term makes the value of the likelihood $L$ essentially zero, as the surface density models give no probability of detecting stars in a zero surface area $dS$. Nonetheless, this zero-term is just a constant scale factor of $L$, so ignoring this term does not affect either the MCMC analysis or the evaluation of classical likelihood ratios and PBFs explained later on in the section, given that this constant term cancels out because it only depends on the number of stars used for the fitting, which is the same for all profiles.}.
	
	For the dSph surface number density profile, we explore the performance of an empirical King profile (\citeauthor{king62} \citeyear{king62}), an exponential profile, a Sérsic profile (\citeauthor{sersic68} \citeyear{sersic68}) and a Plummer model (\citeauthor{plummer11} \citeyear{plummer11}). The surface density of foreground/background contaminants is instead modeled as a bilinear distribution, $f_{\rm cont}=\overline{\rho}\,(1+a\xi+b\eta)$, in order to account for the expected spatial gradient in the density of MW contaminants, due to the relatively large area of the DECam data-set and its location on the sky\footnote{We remind the reader that the classical expression of the King profile is only defined until the tidal radius. Hence, the positive values given by its expression beyond it have to be replaced by zero independently of the fitting process carried out (see an example with nonzero values beyond the tidal radius in Fig.\,5 from \citeauthor{okamoto17} \citeyear{okamoto17}).}.
	
	The likelihood of each structural parameter can then be evaluated through the expression of $\ln L$ (Eq.\,\ref{eq:ln_L}) using a Bayesian Markov Chain Monte Carlo (MCMC) analysis. In this case we used a Bayesian MCMC ensemble sampler with affine invariance of the form from \cite{goodman10}: “The MCMC Hammer” (\citeauthor{foreman-mackey13} \citeyear{foreman-mackey13}); the code we are using\footnote{\url{https://github.com/grinsted/gwmcmc}} was developed at Centre for Ice and Climate (Niels Bohr Institute). This code allows us to sample the entire parameter space without fixing any value and thanks to its affine invariance the performance does not depend on the aspect ratio in probability distributions strongly affected by anisotropy. In the MCMC Hammer we defined 80 walkers, each of them doing approximately $10^4$ steps. The priors for the structural parameters were chosen according to the values given by IH95 and by restricting the domain of each parameter to the region physically possible. 
	
\begin{figure}
		\centering
		\includegraphics[trim={11.45cm 0cm 14.25cm 0cm},clip,width=\hsize]{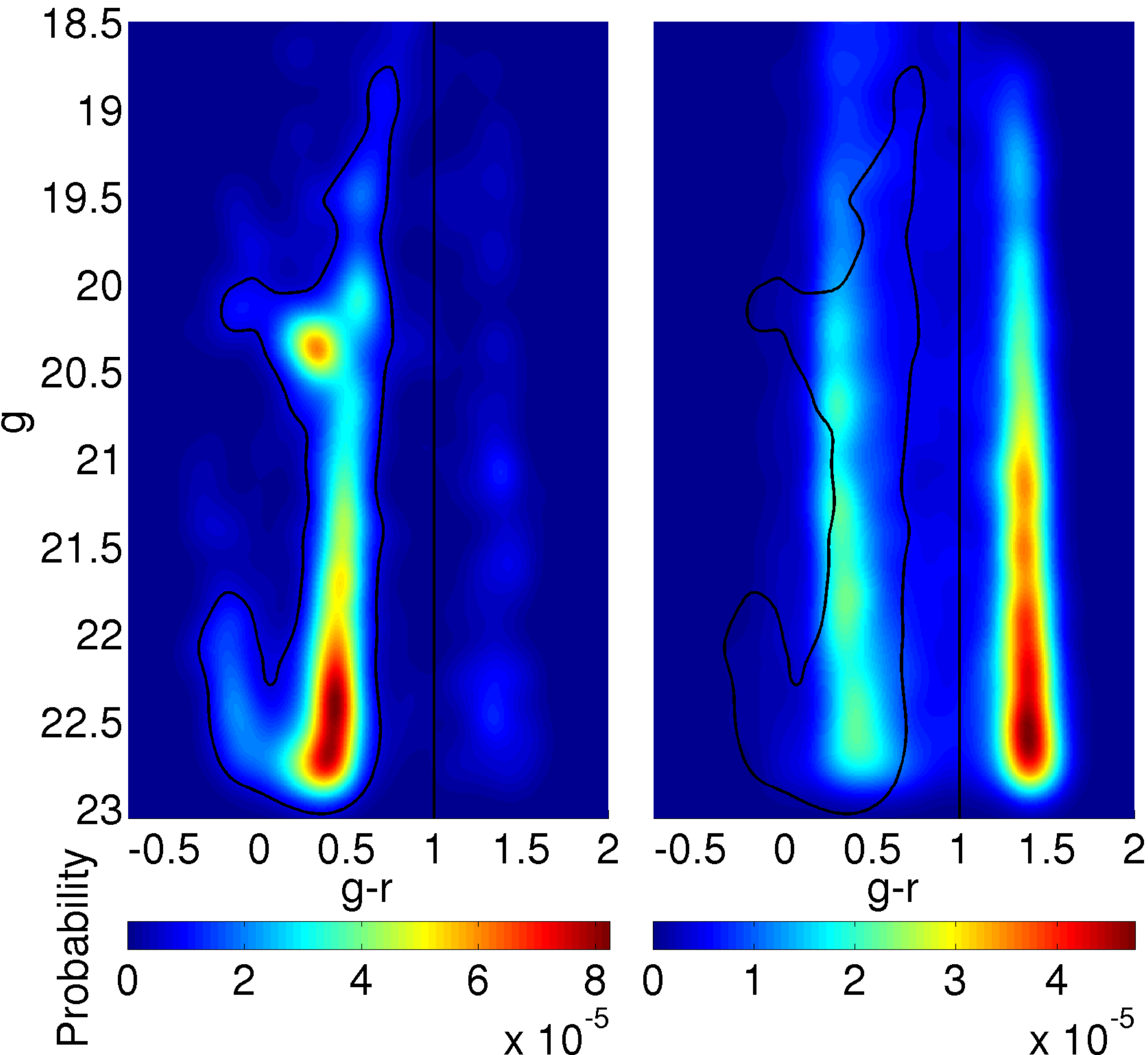}
		\caption{High resolution Hess diagrams from two different spatial regions. Left: central parts of Sextans (MFM signal filter). Right: pointing \#8 (MFM contamination filter). Stars redder than $g-r=1$ (black vertical line) were excluded from the analysis. Black contours: window defined to exclude most contaminants. Hess diagrams are colour-coded according to the probability of finding a star with a given mag and colour.}
		\label{hess}
	\end{figure}

	Due to Sextans' relatively low Galactic latitude, a large number of (mainly foreground) Milky Way stars are expected to contaminate the photometric catalogue, in addition to unresolved background galaxies. We exploit the information contained in the CMD so as to limit the amount of such contaminants. Fig.\,\ref{hess} shows high resolution Hess diagrams, built over a very fine grid in colour and magnitude, which we obtained by first applying a gaussian smoothing dependent on each object's location in accordance to its photometric errors in both bands and then a global gaussian smoothing to fill in the remaining holes caused by the low density of point-like objects on the CMD, with $\sigma_{g-r}\simeq0.06$ mag and $\sigma_{g}\simeq0.1$ mag. The left panel shows the Hess diagram corresponding to the central part of Sextans (i.e. within the central $\sim$10 arcmin), while on the righthand side we show the one corresponding to the displaced pointing (\#8). From now on, unless otherwise stated, we will only consider point-like sources inside the region of the CMD where most of Sextans stellar population is contained, defined as the region within a contour at approximately 10\% the maximum value of the Hess diagram from the central part of the galaxy and excluding objects with $g-r>1$, as they correspond mainly to MW dwarf stars. On the other hand, in order to gain statistics, the $a$ and $b$ parameters in the expression of $f_{\rm cont}$ were derived without using the window in the Hess diagram (Fig.\,\ref{hess}).
	
	For the contaminants, we obtain $\overline{\rho}=0.274\substack{+0.003\\-0.003}$ stars/arcmin$^2$, $a=-1.27\substack{+0.19\\-0.17}$ \%/deg and $b=-4.37\substack{+0.45\\-0.45}$ \%/deg, which produces a trend of increased contamination in the same direction as in the Besan\c con model of the Milky Way stellar populations (\citeauthor{besancon03} \citeyear{besancon03}), as expected, and yields a difference of $\sim$30\% contamination between the edges of our DECam data-set, supporting the appropriateness of allowing for a spatially variable contamination density.
	
\begin{figure*}
		\centering
		\includegraphics[trim={1.5cm 0cm 0cm 0.75cm},clip,width=\hsize]{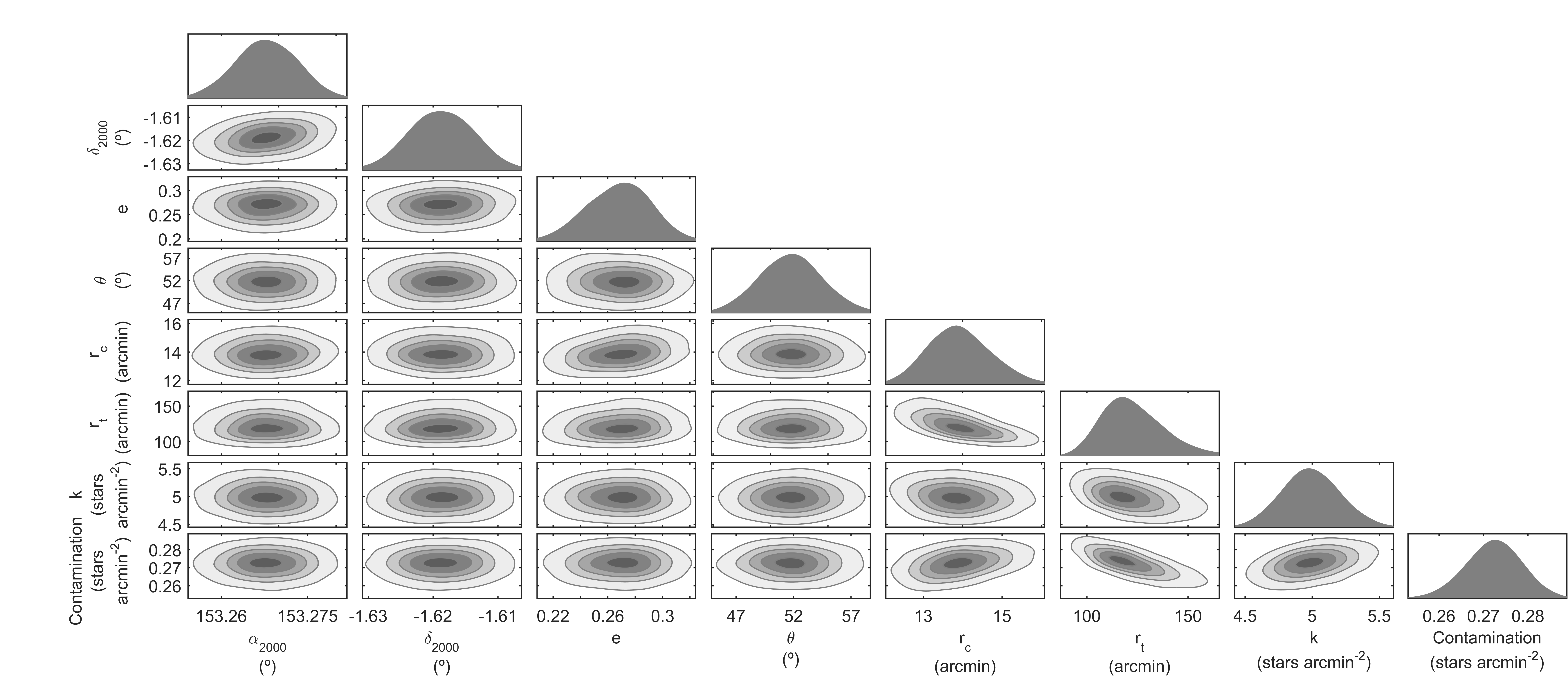}
		\caption{Bayesian posterior distributions of the structural parameters obtained with the MCMC Hammer when modeling the dSph surface number density as an empirical King profile. Contours contain 10\%, 30\%, 50\%, 70\% and 90\% of the points sampled by the MCMC Hammer. From left to right: RA ($\alpha_{2000}$), DEC ($\delta_{2000}$), ellipticity $(e = 1-b/a)$, position angle ($\theta$), core radius ($r_{c}$), tidal radius ($r_{t}$), scale factor (k) and contamination density. Two-dimensional normal distributions aligned with x- and y-axes are indicative of uncorrelated parameters.}
		\label{fig_king_posteriors}
	\end{figure*}

\begin{figure}
		\centering
		\includegraphics[trim={2.5cm 0.5cm 9.75cm 0cm},clip,width=\hsize]{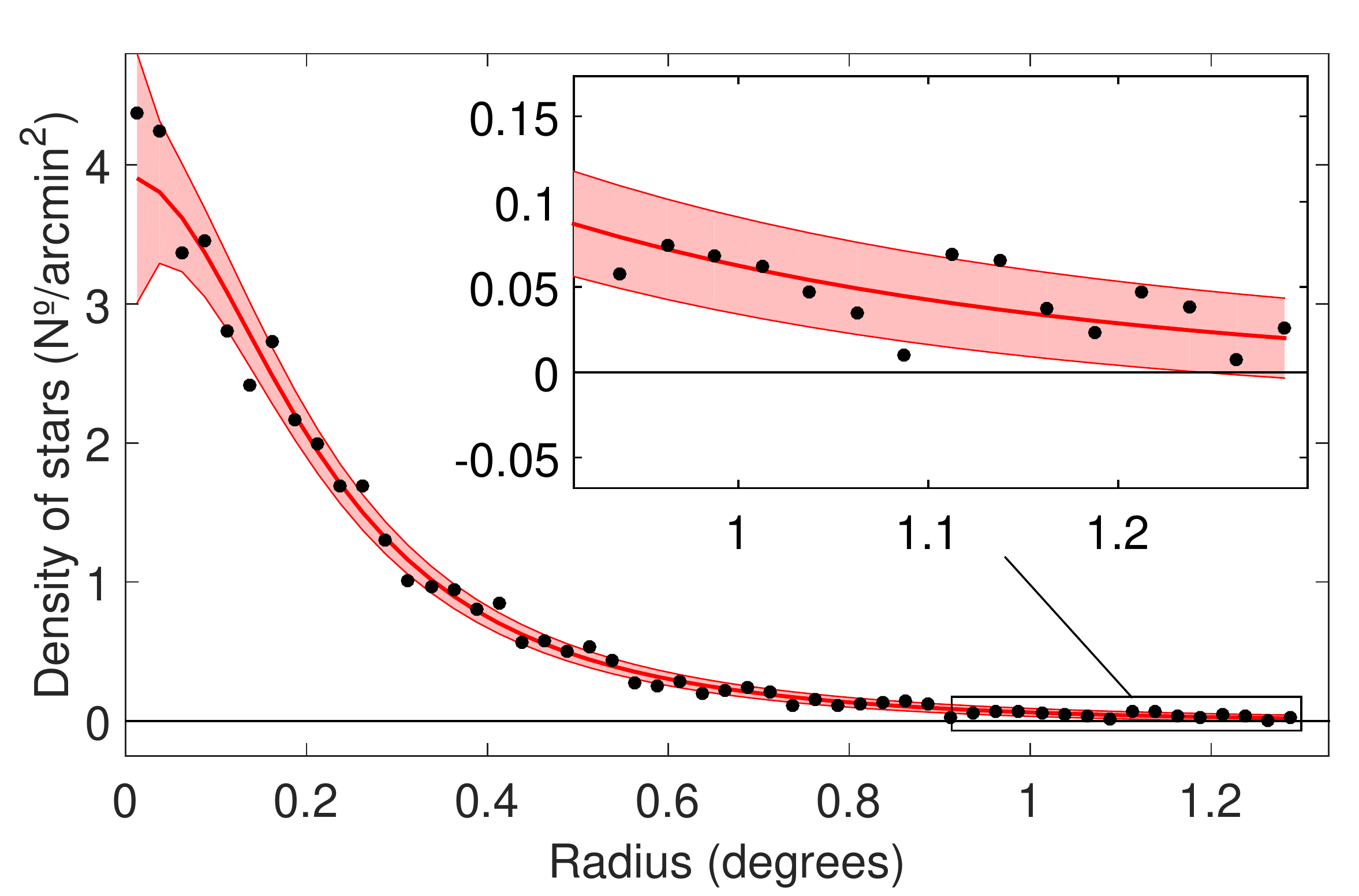} 
		\caption{Contamination subtracted surface number density profile of Sextans stars as a function of the major axis radius (with the external parts zoomed in), overlaid onto the $1\sigma$ confidence interval (red band) of the best-fitting King profile obtained with the MCMC Hammer. The $1\sigma$ confidence interval is computed from the best-fitting model assuming Poisson variances in each elliptical annulus.}
		\label{fig_fit_king}
	\end{figure}

	As an example, Fig.\,\ref{fig_king_posteriors} shows the Bayesian posterior distributions of the parameters obtained when modeling the dSph surface number density as an empirical King profile: we can see that the distributions are nearly Gaussian, hence well behaved, and the parameters well-constrained, with most of them being uncorrelated to each other. The good performance of the best-fitting King model is also evidenced by the agreement with the observed surface number density profile as a function of major axis radius in Fig.\,\ref{fig_fit_king}. The comparison between data and model also shows no significant overdensities of stars with respect to an empirical King profile which could have been interpreted as extra-tidal stars.
	
	{Table\,\ref{structural_parameters} summarizes the results of the procedure. The 2D half-light radius $r_h$ corresponding to the various functional forms were derived using the formulas in \cite{wolf10}. The values and errors quoted are derived from the median and percentiles 15.87-84.13 (corresponding to $\pm1 \sigma$ in case of normal distributions) of the marginalized Bayesian posterior distributions. In Appendix B we tested the performance of our fitting method by applying it to a set of mock galaxies created under different conditions, such as number of stars, galaxy's ellipticity and spatial coverage of the data-set with respect to the galaxy's extent. We find that, for conditions similar to the ones of the photometric sample we are using here, the input parameters for all the generated mock galaxies are within the 1$\sigma$ confidence level; it should however be kept in mind that this approach has implicit the assumption that the models are a good representation of the data. Nonetheless, we can see that the values of the center, ellipticity, position angle and 2D half-light radius we obtain from our DECam catalogue (Tab.\,\ref{structural_parameters}) are stable against the different functional forms for the surface number density profile, within the formal 1$\sigma$ uncertainties.
		
	Table\,\ref{structural_parameters} lists several indicators of the goodness-of-fit, such as the reduced chi-squared $\chi^2_{red}$, the classical ratio of maximum likelihoods, as well as the Posterior Bayes Factor (PBF), in the alternative way described by \cite{aitkin91}. The evidence is classified using the scheme in Table\,\ref{kass_class}. We refer the reader to Appendix C for an extensive explanation on the definition of these goodness-of-fit indicators and their advantages/disadvantages. Here it suffices to say that we consider the PBF as the most reliable discriminant between different profiles, since it does not require any assumption on the data distribution around the model nor spatial binning, and it takes into account the full posterior distribution of parameters values. In order to allow for a fair comparison with the best-fitting models derived in the literature, we also re-perform the MCMC Hammer analysis on our data-set by fitting only the scale factors and contamination densities corresponding to our catalogue, while keeping the rest of the structural parameters fixed to the values published by IH95 and R16, and computing the corresponding values of the goodness-of-fit indicators.
		
	Our determinations of the center, ellipticity and position angle are in good agreement with those by IH95 and R16. On the other hand, our 2D half-light radius is smaller than previous determinations and in very strong disagreement with the half-light radius derived from the Plummer profile of R16. The classical likelihood ratio and PBF favour the King profile and show a very strong evidence (overwhelming in the scale of \cite{aitkin91}) in favour of the set of structural parameters derived in this work over previous determinations. Our photometric catalogue covers a very wide area and reaches deeper than the data from photographic plates on which IH95 work was based on, hence it is reasonable to expect the structural parameters to be better constrained here. On the other hand, the disagreement with the R16 work is unclear (but see Appendix A for some considerations). Here we note that the 2D half-light radius derived from the best-fitting parameters of the R16 King and Plummer model are incompatible with each other, which is not seen either between our four models or in the IH95 best-fitting King vs exponential. Further, we also note that using the structural parameters provided by R16, we could not reproduce the Plummer profile in their Fig. 8. 
		
	In summary, the overall stellar population of Sextans is likely to be more spatially concentrated than previously considered, with a 2D half-light radius of $\sim$22 arcmin, and with a shorter tidal radius, of $\sim$120 arcmin. This approach confirms the King empirical profile as the best representation of the surface number density of Sextans given the DECam catalogue used here, as well as the corresponding values of the centre, ellipticity, position angle (and half-light radius).
		
	In Appendix D we repeat the analysis using the photometric catalogue cut at S/N=5 of the shallowest pointing, i.e. $(g,r)=(24.9,24.9)$, in order to have larger statistics ($\sim$ 440,000 objects vs. $\sim$ 75,000 previously). We relegated this analysis to an appendix because, although some of the fitted structural parameters were significantly different from those in Table\,\ref{structural_parameters} (e.g. $r_{h} = 17\substack{+1\\-1}$ arcmin for the King model), we cannot exclude that the values obtained could be affected by small differences in depth between some pointings and the features corresponding to the out-of-focus regions.
		
	We note that Sextans's dynamical mass within the half-light radius would become 80\% and 60\% of the values given in \cite{walker10} when adopting the 2D half-light radius from the King model $r_h$ we determined from our baseline ``bright'' and S/N=5 catalogues respectively. This would increase the scatter in \cite{walker10} Fig.\,5 in the regime of large r$_h$, but would not deviate considerably from the overall relation.      
		
		\begin{table*}
			\caption[.]{Sextans structural parameters (median values of the marginalized posterior distributions) derived with the MCMC Hammer, plus $\chi^2_{red}$, classical likelihood ratios and Posterior Bayes Factors of the different surface density profiles. Classical likelihood ratios and PFBs are colour-coded according to Table \ref{kass_class}, with blue, magenta and red fonts associated to evidences classified as not worth more than a bare mention, positive evidence and very strong evidence, respectively. The last two columns refers to the best-fitting profiles for which the structural parameters and scale radii were fixed to the values derived by IH95 and R16.}           
			\label{structural_parameters}      
			\centering                          
			\begin{tabular}{c c c c c c c}        
				\hline\hline                 
				\noalign{\smallskip}
				Parameter & Exponential & Sérsic & Plummer & King & IH95 & R16\\    
				\noalign{\smallskip}
				\hline                       
				\noalign{\smallskip}
				$\alpha_{2000}$ (º) & $153.268\substack{+0.006\\-0.006}$ & $153.269\substack{+0.006\\-0.006}$ & $153.268\substack{+0.006\\-0.006}$ & $153.268\substack{+0.006\\-0.006}$ & $153.2625\substack{+0.0005\\-0.0005}$ & $153.277\substack{+0.003\\-0.003}$ \\     
				\noalign{\smallskip}
				$\delta_{2000}$ (º) & $-1.618\substack{+0.006\\-0.006}$ & $-1.618\substack{+0.006\\-0.005}$    & $-1.620\substack{+0.006\\-0.006}$ & $-1.619\substack{+0.006\\-0.005}$ & $-1.6147\substack{+0.0003\\-0.0003}$ & $-1.617\substack{+0.008\\-0.008}$ \\
				\noalign{\smallskip}
				Ellipticity & $0.27\substack{+0.03\\-0.03}$ & $0.27\substack{+0.03\\-0.03}$ & $0.27\substack{+0.03\\-0.03}$ & $0.27\substack{+0.03\\-0.03}$ & $0.35\substack{+0.05\\-0.05}$ & $0.29\substack{+0.03\\-0.03}$ \\
				\noalign{\smallskip}
				Position angle (º) & $52\substack{+3\\-3}$ & $52\substack{+3\\-3}$ & $52\substack{+3\\-3}$ & $52\substack{+3\\-3}$ & $56\substack{+5\\-5}$ & $58\substack{+6\\-6}$\\
				\noalign{\smallskip}
				Sérsic index n & - & $1.03\substack{+0.07\\-0.07}$ & - & - & - & - \\
				\noalign{\smallskip}
				Sérsic factor b(n) & - & $1.6\substack{+0.7\\-0.6}$ & - & - & - & - \\		
				\noalign{\smallskip}
				Exponential $r_{e}$ ($'$) & $12.7\substack{+0.4\\-0.4}$ & - & - & - &  $15.5\substack{+0.1\\-0.1}$ & - \\
				\noalign{\smallskip}
				Plummer $r_{p}$ ($'$) & - & - & $22.8\substack{+0.7\\-0.7}$ & - & - & $35.7\substack{+0.7\\-0.7}$\\
				\noalign{\smallskip}
				Sérsic $r_{s}$ & - & $20\substack{+8\\-8}$ & - & - & - & - \\
				\noalign{\smallskip}
				King $r_{c}$ ($'$) & - & - & - & $13.8\substack{+0.9\\-0.9}$ & $17\substack{+2\\-2}$ & $27\substack{+2\\-2}$ \\
				\noalign{\smallskip}
				King $r_{t}$ ($'$) & - & - & - & $120\substack{+20\\-20}$ & $160\substack{+50\\-50}$ & $83\substack{+8\\-8}$ \\
				\noalign{\smallskip}
				2D Half-light $r_{h}$ ($'$) & $21.4\substack{+0.7\\-0.6}$ & $20\substack{+8\\-8}$ & $22.8\substack{+0.7\\-0.7}$ & $22\substack{+2\\-2}$ & $26\substack{+0.2\\-0.2}$ (Exp.) &  $35.7\substack{+0.7\\-0.7}$ (Plummer) \\
				& & & & & $28\substack{+5\\-5}$ (King) & $24\substack{+2\\-2}$ (King) \\
				\noalign{\smallskip}
				$\chi^2_{red}$ & 0.92 & 1.03 & 0.88 & 1.06 & 1.22 (Exp.) & 5.53 (Plummer) \\ & & & & & 1.57 (King) & 3.15 (King) \\
				\noalign{\smallskip}
				$2\ln{\left(\frac{\textnormal{likelihood}_{1}}{\textnormal{likelihood}_{2}}\right)}$ & \textcolor{blue}{0.3} $\left(\frac{\textnormal{Exp.}}{\textnormal{Sérsic}}\right)$ &\textcolor{magenta}{2.2} $\left(\frac{\textnormal{Sérsic}}{\textnormal{Plummer}}\right)$ & - & \textcolor{magenta}{2.7} $\left(\frac{\textnormal{King}}{\textnormal{Exp.}}\right)$ & \textcolor{red}{-63.1} $\left(\frac{\textnormal{Exp. IH95}}{\textnormal{Plummer}}\right)$ & \textcolor{red}{-336.9} $\left(\frac{\textnormal{Plummer R16}}{\textnormal{Plummer}}\right)$ \\ & & & & & \textcolor{red}{-48.2} $\left(\frac{\textnormal{King IH95}}{\textnormal{Plummer}}\right)$ & \textcolor{red}{-160.7} $\left(\frac{\textnormal{King R16}}{\textnormal{Plummer}}\right)$ \\
				\noalign{\smallskip}
				$2\ln{\textnormal{(PBF)}}$ & \textcolor{blue}{0.8} $\left(\frac{\textnormal{Exp.}}{\textnormal{Sérsic}}\right)$ & \textcolor{blue}{1.8} $\left(\frac{\textnormal{Sérsic}}{\textnormal{Plummer}}\right)$ & - & \textcolor{blue}{1.6} $\left(\frac{\textnormal{King}}{\textnormal{Exp.}}\right)$ & \textcolor{red}{-58.2} $\left(\frac{\textnormal{Exp. IH95}}{\textnormal{Plummer}}\right)$ & \textcolor{red}{-331.9} $\left(\frac{\textnormal{Plummer R16}}{\textnormal{Plummer}}\right)$ \\ & & & & & \textcolor{red}{-32.7} $\left(\frac{\textnormal{King IH95}}{\textnormal{Plummer}}\right)$ & \textcolor{red}{-157.7} $\left(\frac{\textnormal{King R16}}{\textnormal{Plummer}}\right)$ \\
				\noalign{\smallskip}
				\hline             
			\end{tabular}
		\end{table*}
		
		\subsection{Integrated magnitude and central surface brightness}
		
		Here we use our data to revisit Sextans' integrated magnitude and central surface brightness values. Since our data-set does not reach down to several magnitudes below the oldest main sequence turn-off, we supplement it with a synthetic colour magnitude diagram, in order to calculate the flux we are missing from the regions of the CMD we are not sampling.
		
		The synthetic CMD was computed through the algorithm IAC-STAR\footnote{\url{http://iac-star.iac.es}} \citep{iac-star}. We used a star formation history and metallicity law broadly consistent with the observed properties for Sextans stars found by \cite{lee09} and \cite{battaglia11} respectively: the star formation rate was assumed constant for 13.7 to 10 Gyr ago, and null from 10 Gyr ago to the present day, while the metallicity was assumed to range between Z=0.0002 \& 0.0008. We adopted the Teramo stellar evolutionary library (\citeauthor{teramo} \citeyear{teramo}) with the \cite{lejeune97} bolometric correction library, and adopted the default values for the other fields. The synthetic CMD was then shifted to Sextans' distance, adopting a distance modulus (m-M)$_0= 19.67$ (\citeauthor{lee09} \citeyear{lee09}), and cut to match the range in magnitude and colour of our photometric catalogue; the Johnson-Cousin magnitudes were transformed into the SDSS system using the (B-V) $\rightarrow$ (g-B) and (V-R) $\rightarrow$ (r-R) equations by \cite{jordi06}. The surface density profiles fitted to the photometric catalogue were used to obtain the enclosed total number of Sextans stars; from this we calculated a correction ratio, $K$, between the number of stars within $\sim$ 100\% completeness in our photometric catalogue and in the same magnitude range in the synthetic CMD.
		
		Assuming that the synthetic CMD is a fair representation of Sextans' CMD at all magnitudes, the integrated magnitude is:
		\begin{equation}
		V=-2.5\log{\left(K\sum_{i=1}^{n} 10^{-0.4m_{V,i}}\right)}
		\label{eq:mag}
		\end{equation}
		where $m_{V,i}$ is the apparent V-mag of the stars in the synthetic CMD. The error in the integrated magnitude was calculated from the standard deviation of the total number of stars enclosed by the different surface density profiles and the error $\pm0.1$ mag in the distance modulus (derived following \citeauthor{dolphin02} \citeyear{dolphin02}). 
		
		With regard to the central surface brightness we inferred it from the Plummer profile, as it was the most likely in the central pixel of the decontaminated map. We calculated it through equation \ref{eq:mag} as well, but replacing the scale factor $K$ by the ratio between the Plummer central density and the number of synthetic stars in the cut range of the CMD. The error on the central surface brightness was derived from the error in the central density of the best-fitting Plummer profile, obtained via the MCMC, and the error in the distance modulus.
		
		This results in an apparent, absolute magnitude and central surface brightness in V-band: $V=10.73\substack{+0.06\\-0.05}$ mag, $M_V=-8.94\substack{+0.11\\-0.09}$ mag and $\mu_V=27.25\substack{+0.06\\-0.05}$ mag/arcsec$^{-2}$, respectively (these values are already corrected for Galactic extinction). We checked that the adoption of a different stellar library, e.g. the \cite{bertelli94} one, produced values within the derived errors (although we note that the Teramo stellar better reproduces the observed CMD). These values are compatible with those by IH95, $M_V=-9.2\pm0.5$ mag and $\mu_V=27.1\pm0.5$ mag/arcsec$^{-2}$, but with considerably smaller errors.
		
		\section{Surface density maps}
		\label{sec:maps}
		If present, tidal features are likely to have a very low surface density, hence efficient decontamination techniques are needed to enhance their signal over the numerous contaminants, such as ``matched-filtering'' methods (MFM). These essentially exploit the different distributions of the source population and the contaminant population in some combination of observables (e.g. \citeauthor{kepner99} \citeyear{kepner99} and \citeauthor{rockosi02} \citeyear{rockosi02}).
		In the specific case, the Hess diagram of the densest
		regions of the dwarf galaxy, where the ratio contamination/source densities is the lowest, is used to build a ``source'' filter,  defining the shape of the dSph stellar population in the colour-magnitude plane; while a ``contamination filter'' is obtained from a region far enough to be free of galaxy members, i.e. the displaced pointing \#8 (see Fig.\,\ref{hess}). 
		
		Next, we calculated a two-dimensional spatial histogram of the point-like objects in the catalogue. To this histogram we applied the MFM, in the improved form by \cite{mcmonigal14}, which assumes Poisson rather than Gaussian statistics. The observed number of stars $n_i$ per 4 dimensional bin $i$ (spatial: $\xi,\eta$; Hess: magnitude and colour) follows a Poisson distribution of mean $\lambda_i$ = $C_i$ + $\alpha \cdot S_i$, with $\alpha$ being the expected number of dwarf galaxy members in the analysed spatial bin, $C_i$ the contamination filter scaled by the expected number of contaminants in the same bin and $S_i$ the normalized source filter. The probability of the observed number of stars in the 4-dimensional $n_i$ for a given $\alpha$ is: 
		\begin{equation}
		p(n_{i}\,|\,\alpha) = e^{-\lambda_{i}}\frac{\lambda_{i}^{n_{i}}}{n_{i}!}
		\label{eq:prod}
		\end{equation}
		with the probability of $\alpha$ for a given spatial pixel given by the product of the individual probabilities $p(n_{i}\,|\,\alpha)$.
		
		We performed a couple of improvements with respect to the MFM from \cite{mcmonigal14}:
		\begin{itemize}
			\item We substituted the factorial in the Poisson distribution with its continuous version from the gamma function: $n!=\Gamma(n+1)$. Therefore, we do not need to have discrete counts in the bins of the Hess diagrams, allowing us to smooth them to fill in the holes due to the low density of galaxy members and to increment their resolution as much as we wish.
			\item We used a spatially varying contamination model instead of assuming it uniform. In practice, the contamination filter is normalized in the same manner as the source filter but multiplied by the expected number of stars from the contamination in the analysed spatial bin. The derivation of the contamination model is explained in Sect.\,\ref{sec:structural}.  
		\end{itemize}
		
		The only restriction for the size of the spatial bins is to expect at least one contaminant per bin. Otherwise, for smaller bins, in those occupied even by only one contaminant star, $\alpha$ would be always greater than zero, in order to compensate for the fact that the expected number from the contamination filter of the MFM would be always less than one. Furthermore, in the limit of bin size equal to zero, the contamination filter would be null, hence we would not be applying the MFM at all. There is not an upper limit for the spatial bin size, but clearly the wider the bins the larger the number statistics, at the cost of degrading the spatial resolution of the maps. Taking all this into account we settled for a bin size of 4 arcmin per side.
		
		The density map of Sextans' overall stellar population is displayed in the top panel of Fig.\,\ref{mfm_map}, in the form of iso-density contours, after smoothing it with a two-dimensional Gaussian of dispersion $\sigma_{\xi}= \sigma_{\eta}\sim6$ arcmin. We use the results from Sect.~\ref{sec:structural} to determine down to which surface density levels we can confidently trace Sextans's stellar component by using the whole mosaic at once, rather than only pointing \#8: we derive a two-dimensional map of residuals between the MFM decontaminated surface density map of Sextans's stars and the mean surface density predicted by the best-fitting Sersic, King, Plummer and exponential models, as well as a map of the scatter amongst the models. These are visible in the middle and bottom panels of Fig.\,\ref{mfm_map}, respectively, and except for a few over- and under-densities that we will discuss below, show a good agreement between the decontaminated map of Sextans stars and the models, independent on the adopted profile. Based on those maps, we make the assumption that the residuals are mostly artificial fluctuations without any evident spatial trend; we then analyse the frequency distribution of the residuals (inset in middle panel of Fig.\,\ref{mfm_map}) and adopt the 1$\sigma$ confidence interval as the 1$\sigma$ precision of our method; this corresponds to $\sim$0.8 stars/pixel ($\sim$31.8 mag/arcsec$^{-2}$ in V-band), equal to the 1\% of the mean central density from the fitted models. We remind that the requirement of expecting at least one contaminant per cell (and thus the need to have pixels with 4 arcmin per side) plus the two-dimensional Gaussian smoothing of dispersion $\sigma_{\xi}= \sigma_{\eta}\sim6$ arcmin, might hide some small-scale faint substructures whose surface brightness no longer reaches the $\sim$31.8 mag/arcsec$^{-2}$ detection limit in V-band after smoothing them with a kernel of $\sim\sqrt{6^{2}+(4/2)^{2}}\simeq6.3$ arcmin.
		
		At the 2$\sigma$ level, Sextans' density map appears regular, with no significant distortions in the outer parts, except for some overdensities beyond an elliptical radius of $\sim$1 deg. As in R16, there is no detection of large scale tidal structures with a well defined spatial trend (e.g. tidal tails). The over-densities appearing in the displaced pointing might in principle be due to a slightly different photometric zero-point. 
		
		Besides the overdensities visible beyond $\sim$1 deg radius, the two-dimensional map of residuals reveals also other over-dense small clumps as well as under-dense regions, which could not otherwise be appreciated in the surface density map. However, the statistical significance of these features is relatively low, 2$\sigma$/3$\sigma$ away from the background; hence we cannot exclude they are statistical fluctuations. The lack of underdensities beyond the central 1 deg is mainly due to the fact that when we apply the MFM in the form of \cite{mcmonigal14} we discard the possibility of having negative numbers of Sextans stars in any spatial pixel and beyond 1 deg the surface density profiles have low enough values to limit the resulted negative residuals below their 2$\sigma$ detection.
		
		While we do not detect any feature at the location where \cite{walker06} reported the presence a kinematically cold substructure, the one/s reported by \cite{kleyna04} and \cite{battaglia11} are found over a similar spatial region as the central overdensity and could be therefore be related to it. The origin of the other detected over/underdensities is unclear; in particular, to the best of our knowledge, so far there had not been reports of underdensities in the structure of classical dwarf spheroidal galaxies. In all cases the over/underdensities we map are $>$ 8 times away from the scatter amongst the different surface density profiles; this could indicate that most of these detections are real deviations from axi-symmetry of Sextans stellar surface density, independent of the fitted profile. We checked the map of objects classified as extended and found no correspondence between the detected overdensities and background galaxy clusters.
		
		\begin{figure*}
			\centering
			\includegraphics[trim={0cm 0cm 0cm 1.25cm},clip,width=0.9\hsize]{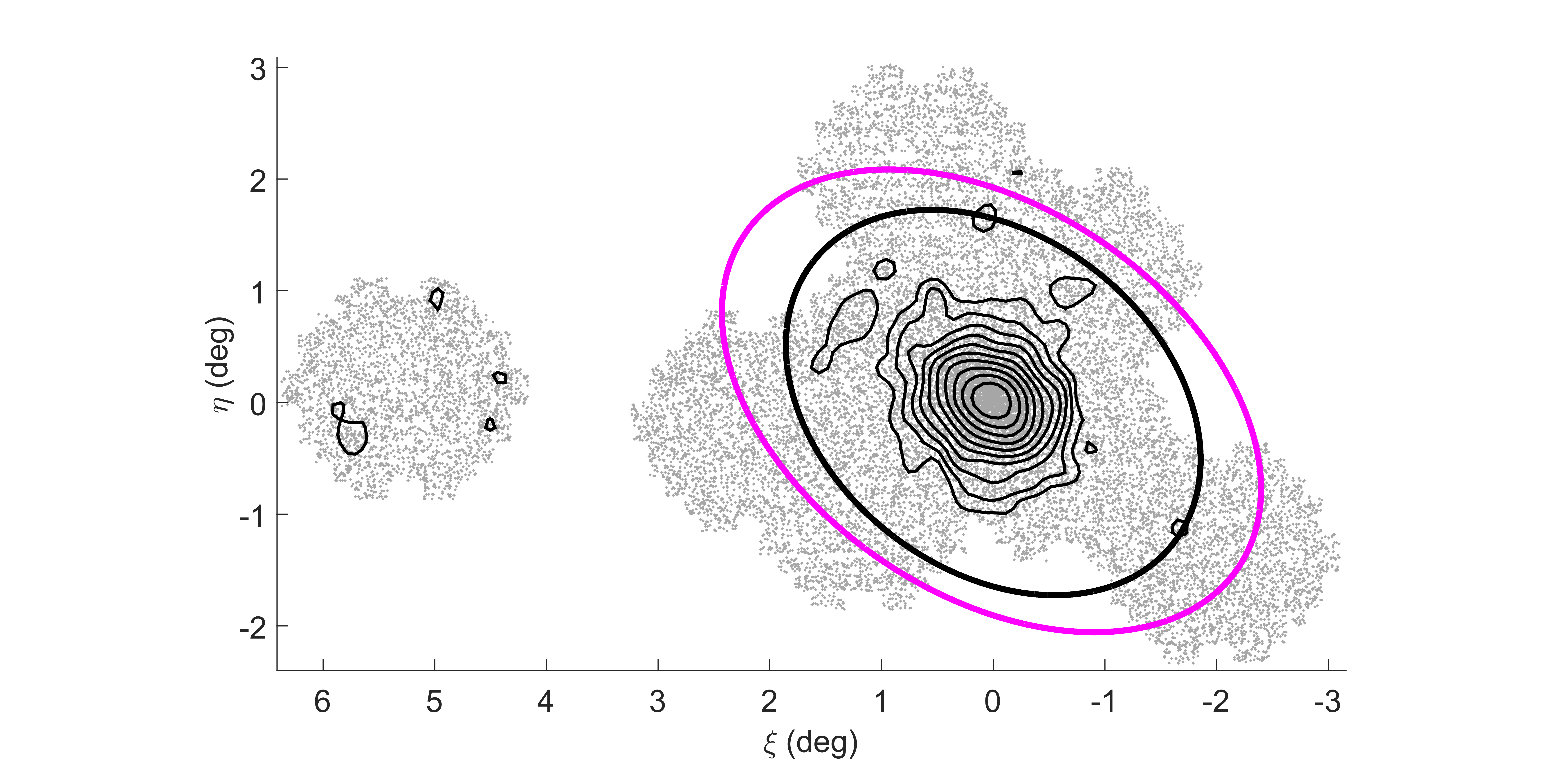}
			\includegraphics[trim={0cm 0cm 0cm 0cm},clip,width=0.905\hsize]{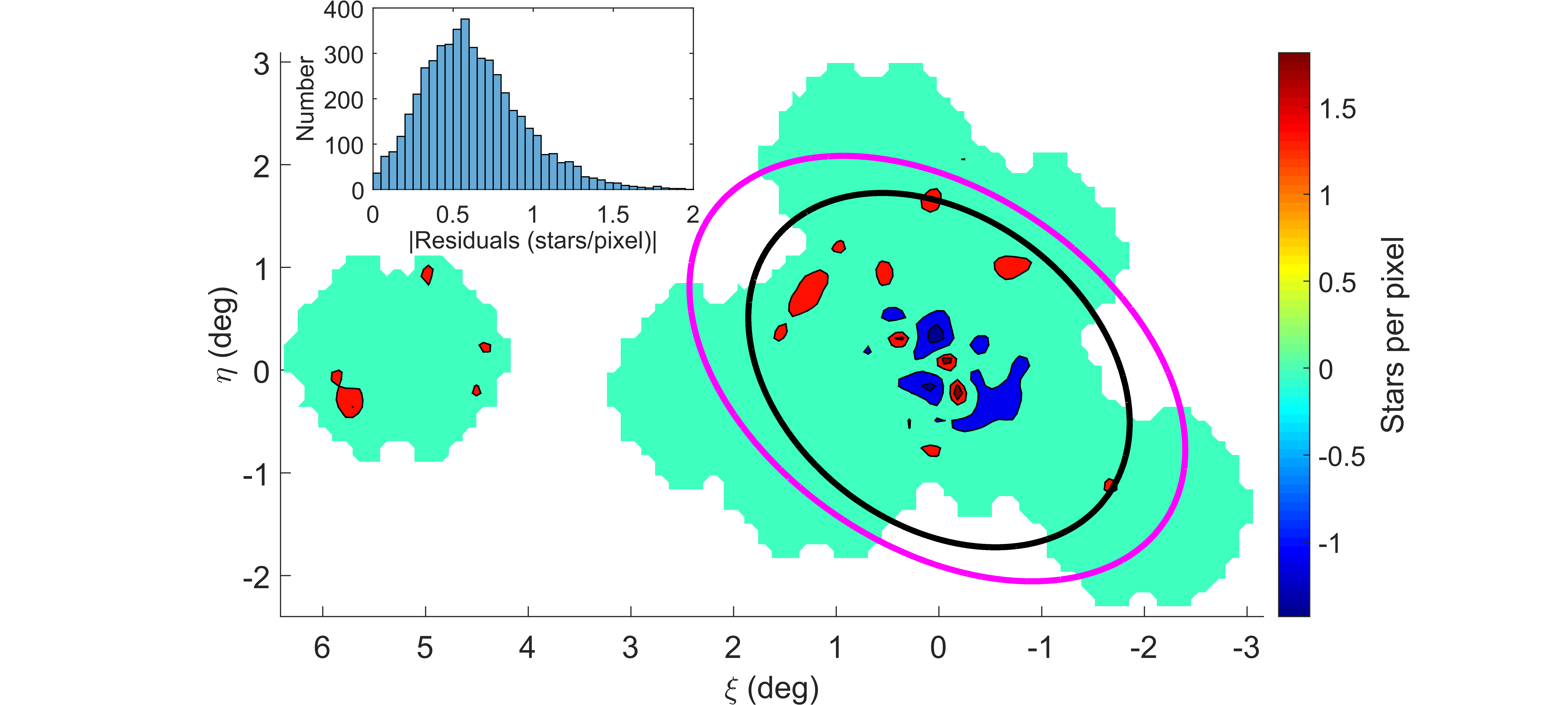}
			\includegraphics[trim={0cm 0cm 0cm 1cm},clip,width=0.905\hsize]{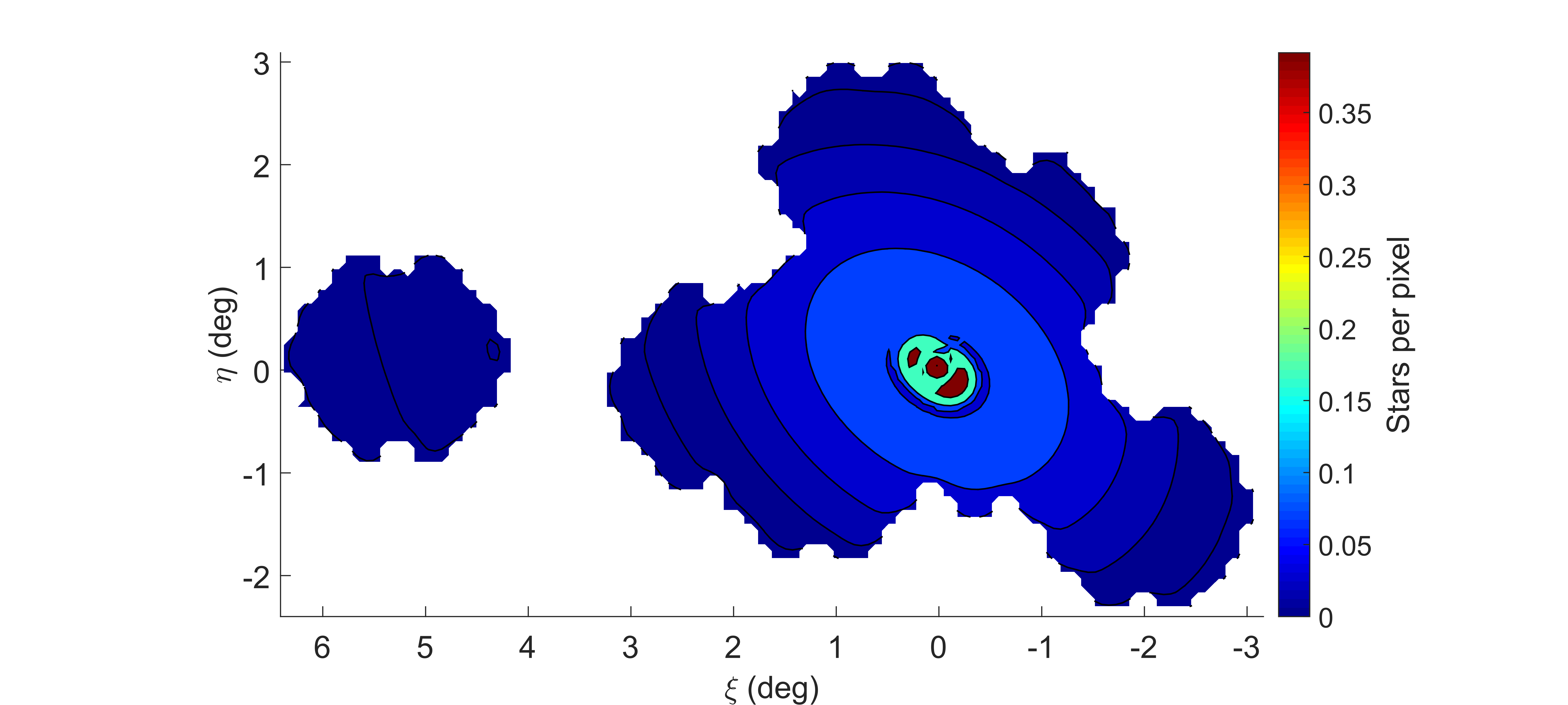}
			\caption{Top: Spatial distribution of stars along the line-of-sight to the Sextans dSph, with overlaid iso-density contours from the MFM surface number density map. The pink and black ellipses show the nominal King tidal radius with parameters from IH95 and this work, respectively. The contours denote exponentially increasing values of surface number density, with the lowest one indicating a 2$\sigma$ detection; the 3$\sigma$ level is approximately the second contour. Middle: Map of residuals between Sextans surface number density map from the MFM analysis and the mean surface density given by the best-fitting King, Sersic, Plummer and Exponential profiles (for the parameters see Table \ref{structural_parameters}), at 2$\sigma$ and 3$\sigma$ detections above and below the mean. The inset shows the frequency distribution of the residuals. Bottom: Standard deviation of residuals between the different fitted profiles; this shows the regions where the residuals depend on a given profile to a greater extent. It can be seen that the over/underdensities hardly depend on the fitted profile, since their surface densities are considerably larger than the scatter between the different profiles.}        
			\label{mfm_map}
		\end{figure*}
			
		\section{Spatial distribution of stars in different evolutionary phases}
		\label{sec:evol_phase}
		
		The analysis of the spatial distribution of stars in different evolutionary phases is a useful tool to study spatial variations of the stellar population mix as a function of age and/or metallicity, also when using relatively shallow photometry. There are multiple examples in the literature of this type of analysis, through which age gradients in Local Group dwarf galaxies have been quantified (e.g. \citeauthor{tolstoy04} \citeyear{tolstoy04}, \citeauthor{battaglia06} \citeyear{battaglia06, battaglia12a, battaglia12b},\citeauthor{mcmonigal14} \citeyear{mcmonigal14}, \citeauthor{bate15} \citeyear{bate15}).
		
		Other possible applications are investigations on the origin of BSs in dSphs, via comparison of their spatial distribution with stars in other evolutionary phases: for example, the significantly more concentrated spatial distribution of BS candidates in Fornax with respect to the overall stellar population likely indicates these are genuine MS stars of young/intermediate age (\citeauthor{mapelli09} \citeyear{mapelli09}), while in Sculptor, Draco and Ursa Minor they are more likely to be actual blue stragglers produced by binaries mass transfer \citep{mapelli07, mapelli09}.
		
		In this section, for the first time we perform a full MFM \& statistical structural analysis of the spatial distribution of Sextans stars in different evolutionary phases, by applying the techniques explained in Sects.\,\ref{sec:structural}, \ref{sec:maps} to Sextans RHB, BHB stars and BSs.
		
		The different stellar populations were isolated through selection windows in the CMD analogous to the one of Fig.\,\ref{hess} but encompassing the RHB, BHB and BS sequence (see pointing \#1 in Fig.\,\ref{cmds}). BSs were also separated into bright ($g<22.3$) and faint ($g>22.3$), as in R16. Due to the lower number statistics when dealing with the sub-sample of stars in different evolutionary phases, we did not calculate the statistical significance of iso-density contours from the distribution of the residuals between the data and the best-fitting model, as it does not reflect the precision of our decontamination method. Instead, we calculate it from the displaced pointing, making the reasonable assumption that it does not contain stars from Sextans. Since our aim here is not to determine which functional form best represents the surface relative density of the various populations, but to quantify relative differences in their spatial distribution, we fit only a Plummer profile, in order to restrict the number of free parameters. As explained in Appendix B, even in the regime of low number statistics of Sextans BHB stars, the MCMC Hammer method provides relatively well-constrained position angle, ellipticity and half-light radii estimates.
		
		Figure \ref{pop_elliptical_annuli} shows that the best-fitting Plummer models give a satisfactory representation of the observed surface density profiles of the various sub-samples. The spatial distributions of the RHB, BHB and bright and faint BSs is shown in Fig.\,\ref{pop_maps}, while the best-fitting structural parameters are summarized in Table \ref{pop_structural_parameters}. Keeping in mind the lower number statistics when analyzing the different evolutionary phases separately, neither the surface density maps or the density profiles show evidence of tidal disturbances for any stellar evolutionary phase.
		
		\begin{figure}
			\centering
			\includegraphics[trim={2.25cm 0cm 9.5cm 0cm},clip,width=\hsize]{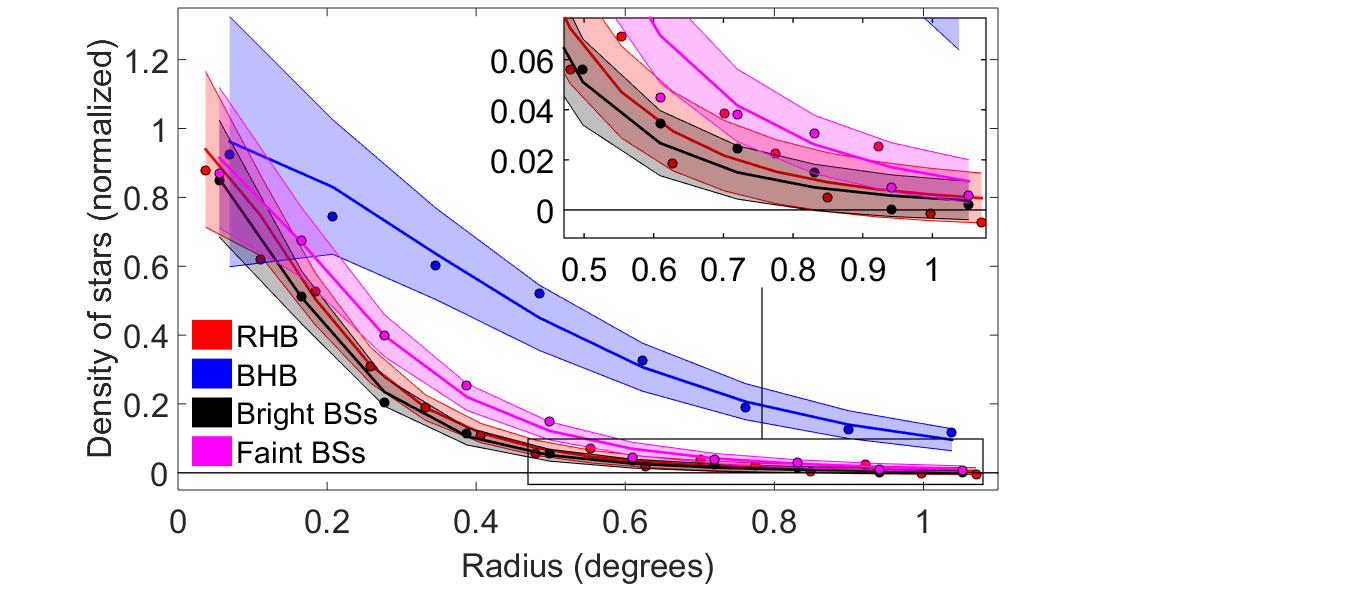}
			\caption{Observed surface density profiles (normalized and contamination subtracted) of RHB, BHB and BSs evolutionary phases as a function of the major axis radius (with the external parts zoomed in). The colour bands show the $1\sigma$ confidence intervals of the best-fitting Plummer profiles obtained with the MCMC Hammer. $1\sigma$ confidence intervals are computed from the fitted models assuming Poisson variances in each elliptical annulus.}
			\label{pop_elliptical_annuli}
		\end{figure}
		
		\begin{figure}
			\centering
			\includegraphics[trim={9cm 0cm 11cm 0cm},clip,width=\hsize]{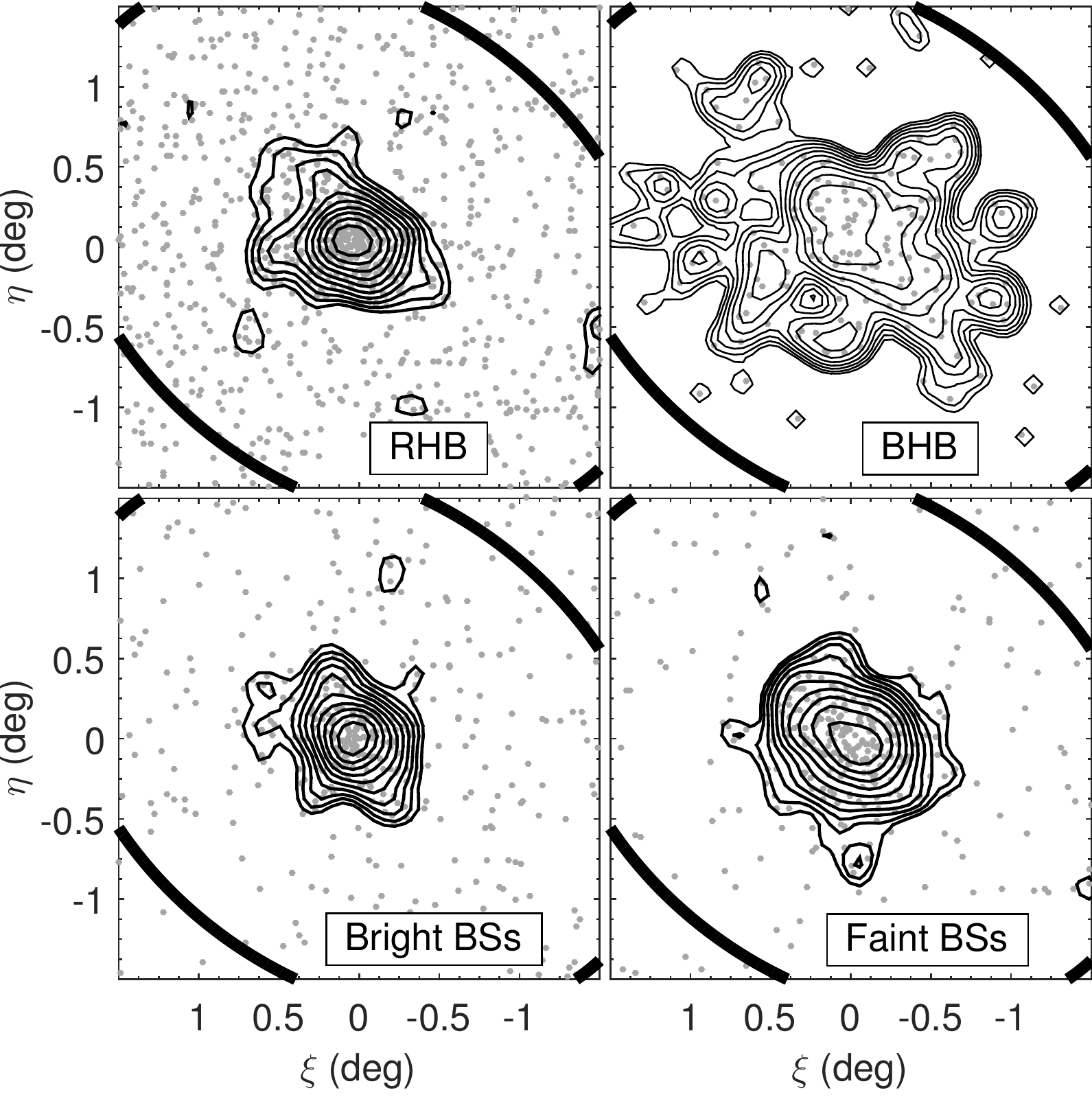}
			\caption{Surface density maps of Sextans RHB, BHB, bright and faint BSs, using the same decontamination techniques from Fig.\,\ref{mfm_map}, with overlaid iso-density contours with exponentially increasing values until the maximum of the measured map (lowest contour at 2$\sigma$ detection from the displaced pointing, with the second one at approximately 3$\sigma$ detection). The ellipse shows the contour of the King profile at its tidal radius derived in this work.}
			\label{pop_maps}
		\end{figure}
		
		\begin{table*}
			\caption[.]{Structural parameters (median values of the marginalized posterior distributions) for Sextans stars in the RHB, BHB and BSs evolutionary phases, derived with the MCMC Hammer when modeling the surface number density of Sextans stellar populations as a Plummer profile; for completeness, we list also the $\chi^2_{red}$ of the best-fitting model and the number of stars analyzed.}           
			\label{pop_structural_parameters}      
			\centering                          
			\begin{tabular}{c c c c c}        
				\hline\hline                 
				\noalign{\smallskip}
				Parameter & RHB & BHB & Bright BSs & Faint BSs\\    
				\noalign{\smallskip}
				\hline                        
				\noalign{\smallskip}
				$\alpha_{2000}$ (º) & $153.29\substack{+0.02\\-0.02}$ & $153.21\substack{+0.05\\-0.05}$ & $153.28\substack{+0.02\\-0.02}$ & $153.25\substack{+0.03\\-0.03}$ \\
				\noalign{\smallskip}
				$\delta_{2000}$ (º) & $-1.60\substack{+0.02\\-0.02}$ & $-1.60\substack{+0.05\\-0.05}$ & $-1.63\substack{+0.02\\-0.02}$ & $-1.64\substack{+0.02\\-0.02}$ \\
				\noalign{\smallskip}
				Ellipticity & $0.29\substack{+0.08\\-0.09}$ & $0.2\substack{+0.2\\-0.2}$ & $0.2\substack{+0.2\\-0.2}$ & $0.25\substack{+0.08\\-0.1}$ \\
				\noalign{\smallskip}
				Position angle (º) & $60\substack{+10\\-10}$ & $60\substack{+20\\-40}$ & $40\substack{+30\\-20}$ & $50\substack{+20\\-20}$ \\
				\noalign{\smallskip}
				2D Half-light $r_{h}$ ($'$) & $17\substack{+2\\-2}$ & $42\substack{+7\\-7}$ & $16\substack{+3\\-3}$ & $22\substack{+3\\-3}$ \\
				\noalign{\smallskip}
				$\chi^2_{red}$ & 1.12 & 1.27 & 0.96 & 1.20 \\
				\noalign{\smallskip}
				N$º$ stars analyzed & 2022 & 212 & 820 & 569 \\
				\noalign{\smallskip}
				\hline              
			\end{tabular}
		\end{table*}
		
		\subsection{Compatibility between the spatial distributions}
		
		In order to statistically analyse the compatibility between the spatial distributions of stars in the different evolutionary phases, we obtained their radial cumulative distribution functions (CDFs)\footnote{This was done using the elliptical radius derived adopting the center, $\epsilon$ and $\theta$ obtained when fitting a King profile (see Table \ref{structural_parameters}). and performed Kolmogorov-Smirnov (K-S) tests between all the possible pairs.} This kind of analysis was already done by \cite{lee03} out to a major axis radius of $\sim$ 25 arcmin - i.e. approximately the half-light radius. Our data-set covers the galaxy out to its nominal King tidal radius (120 arcmin in this work) and beyond, which can lead to different results due to possible changes in the spatial distributions beyond the half-light radius. Furthermore, the effect of contamination on the radial CDFs has been neglected until now (\citeauthor{lee03} \citeyear{lee03}, R16), while it can have a significant impact on the CDFs' shapes and therefore on the results from the K-S tests. Hence, we decided to decontaminate all radial CDFs before performing any analysis.
		
		We use a decontamination method that removes the need for having empirical CDFs evaluated at the same radii or the loss of spatial resolution when using elliptical annuli to deal with this issue. The total unnormalized CDF ($F_{tot} (r)$) of each evolutionary phase is a sum of the unnormalized CDFs of both galaxy members ($F_{gal} (r)$) and contaminants ($F_{cont} (r)$). The decontaminated CDF we are looking for is thus: $F_{gal} (r)$ = $F_{tot} (r)$ - $F_{cont} (r)$, in absolute counts. $F_{tot} (r)$ is directly measurable for each evolutionary phase. As for $F_{cont} (r)$, we first derive its shape - i.e. $F_{cont} (r)$ normalized. Assuming that the CMD of the contamination is constant across the whole field-of-view, the shape of $F_{cont} (r)$ is the same for the contaminants in all evolutionary phases and we infer it by measuring the empirical CDF of objects redder than $g-r=1$ (black vertical line in Fig.\,\ref{hess}). We then calculate the number of contaminants expected in each evolutionary phase, by fitting the bilinear distribution $f_{\rm cont}=\overline{\rho}\,(1+a\xi+b\eta)$ to the spatial distribution of these red contaminants through the MCMC Hammer. We can then calculate the ratios between our newly fitted central density ($\overline{\rho}$) and the ones previously obtained for the contamination in each evolutionary phase when we fitted their structural parameters. These ratios reflect the proportion of contaminants expected in each evolutionary phase with respect to the number of contaminants with $g-r>1$. Thus, to obtain the expected radial distribution of contaminants in an evolutionary phase, we uniformly resample the radii of the objects redder than $g-r=1$ according to its corresponding ratio, to later decontaminate the original sample by removing the stars with the closest radius to that expected for the contaminants.
		
		The decontaminated CDFs are shown in Fig.\,\ref{Fs} per evolutionary phase and for the whole Sextans population; Tab.\,\ref{KS} presents the $p$-values resulting from the K-S tests carried out over all the possible combinations of the CDFs analyzed. The $p$-values are reasonably accurate for sample sizes $n_{1}$, $n_{2}$ such that $(n_{1} \cdot n_{2})/(n_{1}+n_{2})>4$ (\citeauthor{smirnov39} \citeyear{smirnov39}), with our samples satisfying by far this requirement.
		
		\begin{figure*}
			\centering
			\includegraphics[trim={0cm 0cm 0.25cm 0cm},clip,width=\hsize]{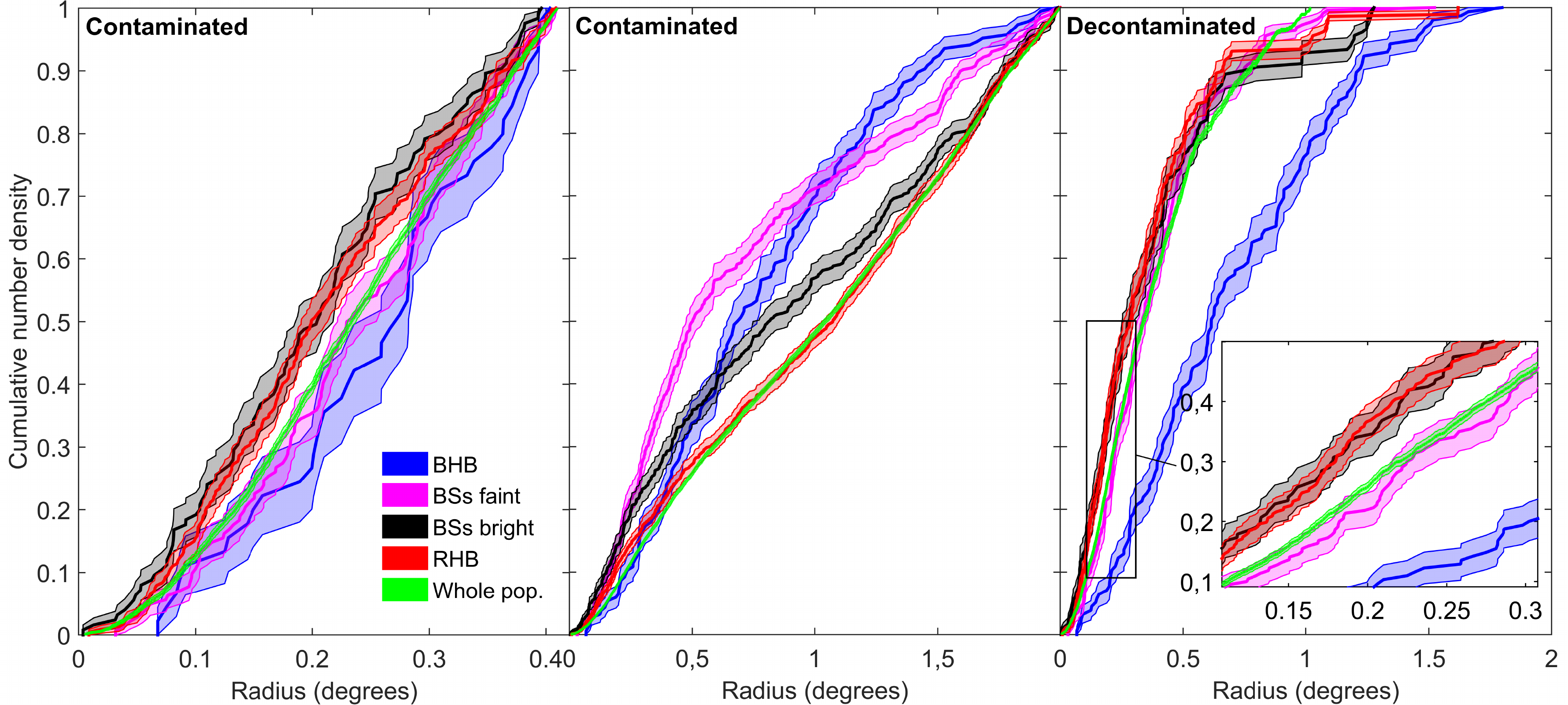}
			\caption{Radial CDFs of the evolutionary phases studied in Sect.\,\ref{sec:evol_phase} and the whole population of Sextans. The colour bands show the $1\sigma$ confidence intervals. Left: Within the major axis radius reached by \cite{lee03}. Although this plot is contaminated, the decontaminated version is rather similar, as it does not reach the external regions of Sextans where the impact of contaminants is higher. Middle: Within the tidal radius derived in this work, with contaminants included. Right: Within the tidal radius derived in this work, with contaminants extracted and an inset magnifying the internal region where all evolutionary phases are better separated.}
			\label{Fs}
		\end{figure*}
		
		\begin{table*}
			\caption[.]{Results from the K-S tests between the radial distributions of the different evolutionary phases (all possible combinations). Column \#2 is just for comparison with results from \cite{lee03}, where contamination is included and the coverage is limited to a major axis radius of $\sim$ 25 arcmin. Columns \#3 and \#4 show the different $p$-values obtained when contamination is included or subtracted, respectively.}           
			\label{KS}      
			\centering                          
			\begin{tabular}{c c c c}        
				\hline\hline                 
				\noalign{\smallskip}
				& & K-S $p$-value (\%) & \\
				\noalign{\smallskip}
				Evolutionary phase & Cont. ($r<25$') & Cont. ($r<120$') & Decont. ($r<120$') \\
				\noalign{\smallskip}
				\hline                        
				\noalign{\smallskip}
				RHB - Whole pop. & 1.2 & 26.4 & $6.8\cdot10^{-3}$ \\
				\noalign{\smallskip}
				BHB - Whole pop. & 21.1 & $7.4\cdot10^{-8}$ & $1.5\cdot10^{-15}$ \\
				\noalign{\smallskip}
				Bright BSs - Whole pop. & 1.5 & 0.019 & 1.2 \\
				\noalign{\smallskip}
				Faint BSs - Whole pop. & 73.4 & $1.8\cdot10^{-19}$ & 31.8 \\
				\noalign{\smallskip}
				RHB - BHB & 1.1 & $8.5\cdot10^{-9}$ & $8.9\cdot10^{-16}$ \\
				\noalign{\smallskip}
				Bright BSs - Faint BSs & 4.0 & $1.3\cdot10^{-3}$ & 2.2 \\
				\noalign{\smallskip}
				RHB - Bright BSs & 93.7 & 2.6 & 31.6 \\
				\noalign{\smallskip}
				BHB - Faint BSs & 54.0 & 0.042 &  $2.6\cdot10^{-11}$ \\
				\noalign{\smallskip}
				RHB - Faint BSs & 2.5 & $3.0\cdot10^{-14}$ & 0.088 \\
				\noalign{\smallskip}
				BHB - Bright BSs & 0.81 & $8.8\cdot10^{-3}$ &  $3.9\cdot10^{-11}$ \\
				\noalign{\smallskip}	
				\hline               
			\end{tabular}
		\end{table*}
		
	When we reproduce the limitations in \cite{lee03} by restricting the coverage until a major axis radius of $\sim$ 25 arcmin and without decontaminating the samples (left panel Fig.\,\ref{Fs} and left column in Tab.\,\ref{KS}), our results agree very well with theirs: with a significance level $\alpha=0.05$, the populations that remain spatially compatible are the RHB with the bright BSs, the BHB with the faint BSs and the last two with the whole population. The results hold also when decontaminating the samples; hence the impact of the contamination is not an issue over these spatial scales. 
		
	The other panels of Fig.\,\ref{Fs} (and columns in Tab.\,\ref{KS}) use CDFs derived out to 120 arcmin, and illustrate the effect of neglecting contamination when comparing the CDFs over this much larger radial extent: except for the BHB, all CDFs are strongly affected, leading to unrealistic shapes and wrong $p$-values derived with K-S tests. When neglecting contamination, the only populations spatially compatible are the RHB stars and the overall stellar population. This is very different from the results obtained when decontaminating the CDFs: the spatial distribution of RHB stars is incompatible with the one of the overall Sextans stellar population; on the other hand, the RHB stars and bright BSs have compatible spatial distributions, and so do the faint BSs with Sextans overall stellar population. By extending the coverage from 25 to 120 arcmin, we have also discarded the possibility that BHB and faint BSs have the same spatial distribution. These results are consistent both numerically (Table\,\ref{KS}) and graphically (confidence intervals in Fig.\,\ref{pop_elliptical_annuli} and \ref{Fs}).}
	
	K-S tests from R16 yielded a $p$-value of 38\% for BHB with BSs, and 99.8\% for bright BSs with faint BSs, both incompatible with our results. It is not clear if R16 performed the K-S tests with the CDFs evaluated with different structural parameters for each evolutionary path or not. This point would invalidate the results from their K-S tests, as these evaluate the compatibility of the different spatial distributions by reducing their comparison to the radial dimension, assuming that this is the only variable reflecting the differences between the distributions. For example, two populations with very different centres or position angles that have clearly distinct spatial distributions can share the same radial CDF if this is separately evaluated for each population using their individual structural parameters. Nevertheless, even in the case that R16 used the same structural parameters for all evolutionary paths, we have proven that the effect of the contamination is not negligible at all with these large spatial coverages, therefore having an undeniable contribution to the discrepancy between our results and those in R16.
	
	\subsection{RHB and BHB}
	
	Our analysis shows that	the BHB stars cover a considerably larger area than the other analysed stellar populations, with a statistically significant measurement of a half-light radius more than twice that of the RHB and much larger than the overall Sextans stellar population.
	
	The finding of RHB stars being more centrally concentrated than BHB stars is in agreement with the early results by \cite{bellazzini01} based on the variation of the relative number of RHB and BHB stars in two spatial bins, as well as with \cite{lee03} and \cite{okamoto17}, based on the CDF of RHB/BHB stars and the color distribution of HB stars as a function of major axis radius. By application of the K-S test, we find an almost zero probability of RHB stars being extracted from the same spatial distribution as BHB stars. It appears highly likely that the differences between the RHB and BHB spatial distributions are the consequence of the age (\citeauthor{okamoto17} \citeyear{okamoto17}) and metallicity (\citeauthor{battaglia11} \citeyear{battaglia11}) gradient detected in the overall old stellar component of Sextans, similar to the case of the Sculptor dSph (\citeauthor{tolstoy04} \citeyear{tolstoy04}, \citeauthor{martinez-vazquez15} \citeyear{martinez-vazquez15}). This was previously observed by \cite{bellazzini01} and \cite{lee09}, who already associated RHB and BHB stars to the more metal-rich/younger (more centrally concentrated) and more metal-poor/older (more spatially extended) populations respectively.
	
	While the BHB spatial distribution shows much irregularities (Fig.\,\ref{pop_maps}), we have verified that this is likely due to small number statistics; indeed by extracting sub-samples of N$_{\rm BHB}$ stars from the more populated BSs and RHB populations, their regular spatial distribution would, artificially appear, very distorted.
	
	\subsection{Candidate Blue Straggler Stars}
	From their analysis of Sextans's central regions, \citet{lee03} noticed that, when separating candidate BSs into a ``bright'' and ``faint'' sample through a magnitude cut, the former had a more centrally concentrated spatial distribution than the latter. The probability of being extracted from a similar spatial distribution was low, however this latter point was questioned by R16.
	
	In our analysis we confirm the early findings by \cite{lee03} that ``bright'' BSs and ``faint'' BSs have a low probability of being extracted from a similar spatial distribution ($p$-value = 2.2\% from the K-S test), and we derive a half-light radius of $16\substack{+3\\-3}$ and $22\substack{+3\\-3}$ arcmin for ``bright'' and ``faint'' BSs, respectively.        
	
	In the case of globular clusters a possible interpretation for this feature would be that BSs created via stellar collisions (\citeauthor{hills76} \citeyear{hills76}) mainly correspond to the bright ones, and those evolving from primordial binaries to the faint ones (\citeauthor{mccrea64} \citeyear{mccrea64}). One reason for this is that collisions are more probable where the stellar density is large, such as in the inner regions of dense clusters, where mass segregation is a dominant process; hence collisions of more massive stars would produce bluer and brighter BSs (\citeauthor{bailyn_pinsonnealut95} \citeyear{bailyn_pinsonnealut95}; \citeauthor{bailyn95} \citeyear{bailyn95}). However, as argued by \citet{lee03}, in dSph galaxies the central densities are so low that BSs created via stellar collisions are highly unlikely (\citeauthor{mapelli07} \citeyear{mapelli07}; \citeauthor{momany07} \citeyear{momany07}). An intriguing hypothesis was proposed by \cite{kleyna04}, who argued that ``if a significant fraction of Sextans BSs were formed in a star cluster which subsequently disrupted near the centre of the galaxy, mass segregation within the cluster would ensure that the most massive (brighter) BSs would be the last to be tidally removed from the cluster and hence would have a more concentrated spatial distribution''. 
	
	Does the different spatial distribution of ``bright'' and ``faint'' Sextans BSs need to be explained by invoking the disruption of a globular cluster that spiralled in the central regions? Or can they be explained as the result of mass transfer from primordial binaries?
	
	In our analysis, the comparison of the cumulative CDFs shows that the ``bright'' BSs have a spatial distribution compatible with being extracted from a similar one than Sextans' RHB stars, whilst the ``faint'' ones are compatible with the overall Sextans stellar population. If Sextans BSs are the result of mass transfer from primordial binaries, one would expect them to show a similar spatial distribution as the overall stellar population, and to possibly reflect the age/metallicity gradient traced by other stellar populations if one is able to associate parts of the BS sequence to stellar populations in other evolutionary phases. 
	
	\citet{lee03} (see their Fig.\,19) showed that 2-6 Gyr old isochrones encompass the range of magnitudes and colours of Sextans BS sequence, with the bluer/brighter BSs being reproduced by the youngest isochrones, without much contamination from the older ones; on the other hand, over the CMD selection of ``faint'' BSs, the isochrones overlap significantly in magnitude and colour. While in the case of BSs the age range of the isochrones is not representative of the stars actual age, this shows that the bluer/brighter BSs (corresponding to the ``bright BS'' selection) can be thought of as slightly more massive on average than the fainter/redder ones (i.e. the ``faint BS'' selection); and that the ``faint'' selection hosts of a mix of the stellar population that gave rise to the BSs, rather than only the less massive end. This could explain why the spatial distribution of the ``bright'' BSs is compatible with the RHB distribution, while the ``faint'' BSs appear to trace Sextans overall stellar population.
	
	Overall, while we cannot discard the possibility that some of Sextans BSs come from a disrupted stellar cluster, the above results do not make this hypothesis compelling to explain the bulk of Sextans BSs. 
	
	Another possibility concerning the nature of Sextans BSs is that they are actual main-sequence stars of intermediate-age ($\sim$2-6 Gyr old). If this were the case, as a consequence of the age gradient detected in Sextans, one would expect these stars to be the most spatially concentrated ones, as they would be the youngest ones of the stellar populations analysed. Our analysis shows that this is not the case, hence we deem it unlikely that Sextans BSs are genuine intermediate-age ($\sim$2-6 Gyr old) main-sequence stars.        
	
	\section{Spectroscopic analysis}	
	\label{sec:spectroscopy}
	
	We use our revised structural parameters of the Sextans dSph to update the membership probabilities of the spectroscopic samples by \cite{walker09} and \cite{battaglia11} (hereafter W09 and B11).
	
	To this aim, we followed the ``expectation maximization" (EM) technique outlined in W09, but with some modification, as explained below.
	
	For the catalogue of W09 the measurements we used are the heliocentric velocity and the pseudoequivalent width of the Mg-triplet absorption feature, while for that of B11 we used the heliocentric velocity, the equivalent width of the Mg {\sc i} line at 8806.8 \AA\,and the metallicity ([Fe/H]). The information on the projected distance from the galactic centre was used for both catalogues. In B11 an hard-cut was used to separate members from nonmembers; here instead we assume a Gaussian distribution for the probability distribution functions of both members and nonmembers in metallicity and Mg {\sc i} line measurements, and the expressions given in W09 for the velocity.
	
	The modifications we apply to W09 EM algorithm are the following:
	\begin{itemize}
		\item In W09 the likelihood and the membership probabilities were calculated separately, in an iterative way, until convergence. Here we inserted their Eq.\,3 into their Eq.\,2 and maximized this expression at once through the MCMC Hammer.
		\item In W09 the mean and variance of the probability distribution of the measurements (e.g. the velocity or metallicity distribution) for members were considered as constant across the whole field-of-view. We allow for radial variations of these quantities by separately running the MCMC Hammer over elliptical annuli drawn from the King profile fitted in Sect.\,\ref{sec:structural}; this takes into account for example the presence of a metallicity gradient.
		\item W09 estimate the a priori probability of membership as a function of radius via monotonic regression; we preferred to derive it from the mean surface density profile fitted in Sect.\,\ref{sec:structural}. We consider it gives us a better description of how the membership probability decays as a function of radius while at the same time not depending on a unique formula for the adopted surface density profile.
	\end{itemize}
	
	\begin{figure*}
		\centering
		\includegraphics[trim={0.5cm 0.75cm 1cm 0.5cm},clip,width=0.8\hsize]{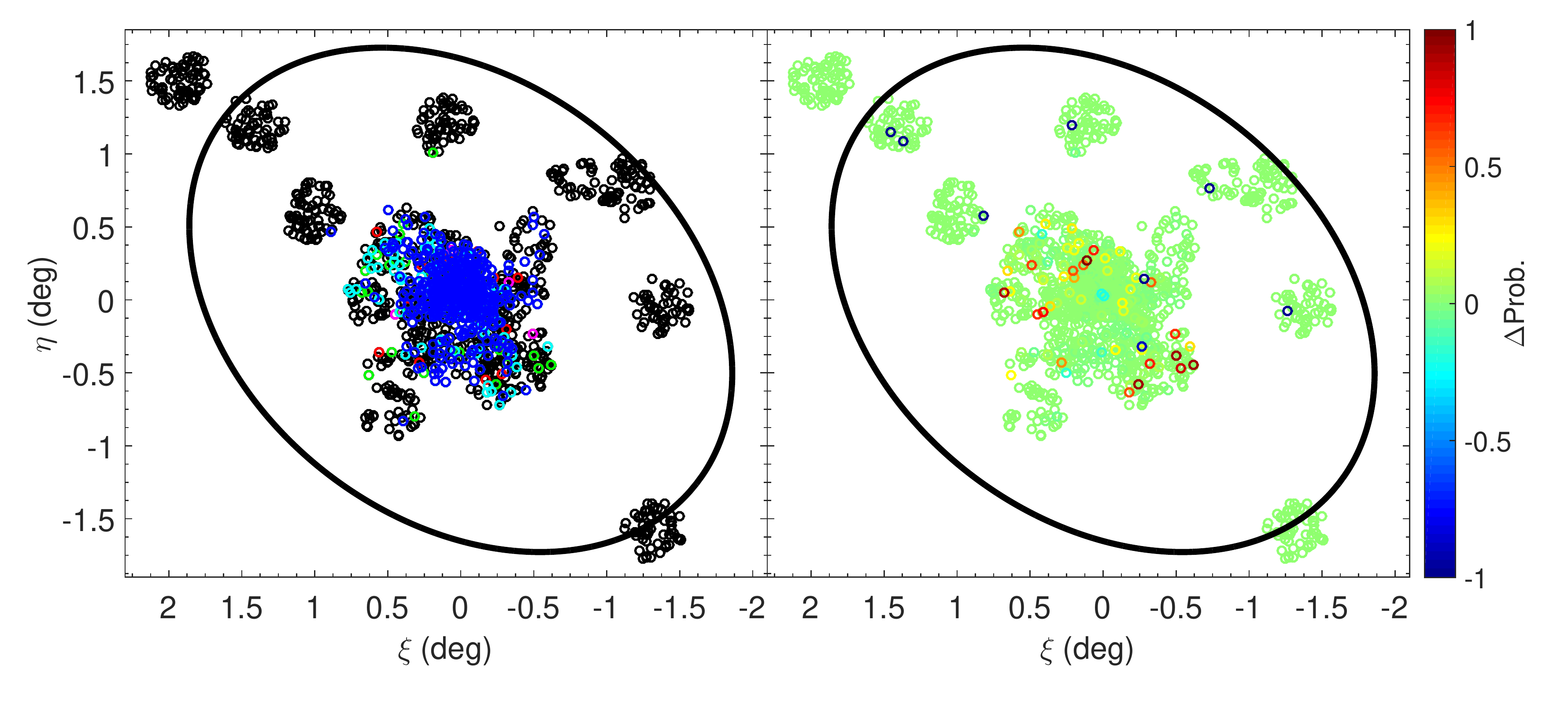}
		\caption{Location of stars along the line-of-sight to Sextans from the spectroscopic samples of W09 and B11. Left: Colour-coded using the updated membership probabilities from this work. The colour-coding is as in W09, with black, red, magenta, green, cyan and blue markers denoting $P_i$ $\geq$ 0.00, > 0.01, > 0.50, > 0.68, > 0.95, > 0.99 respectively. Right: Colour-coded using the difference in the membership probabilities here derived and in the previous works ($\Delta$Prob. = P\textsubscript{new} $-$ P\textsubscript{prev}). Black ellipses show the nominal King tidal radius with parameters from this work. For these plots, in the case of the overlapping stars we opt to use the probabilities derived from the data-set of B11, as when evaluating their probabilities we can rely on three spectroscopic sources of information (v\textsubscript{hel}, EW Mg and [Fe/H]), rather than two as in the case of the W09 data-set (v\textsubscript{hel} and $\Sigma$Mg). Further, we consider that the metallicity is a more reliable discriminant compared to just the Mg-triplet absorption feature from W09, as it depends on three parameters measured by B11: EW Mg of the star and apparent magnitudes in V band of both the star and the horizontal branch of the galaxy.}
		\label{prob_spec}
	\end{figure*}
	
	Table \ref{spectroscopic_catalogue} shows the spectroscopic catalogues with the updated membership probabilities. Excluding from the FLAMES data-set the stars that did not meet the quality requirements established in B11, there are 1595 entries, from which the EM algorithm detected a total of 629 members by summing all their membership probabilities. Of these, 141 stars overlap between the W09 and B11 catalogues, with their positions agreeing within $\sim$1 arcsec, and their
	membership probabilities were calculated for each of the two catalogues separately. The agreement with the previous determinations is good (Fig.\,\ref{prob_spec}); the few discrepant values are likely due to the fact that we are allowing for radial gradients of the properties and that we adopt the average density profile as a prior on membership probability as a function of radius.	
	
	\begin{table*}
		\footnotesize
		\caption[.]{Sample from the combined W09 and B11 spectroscopic catalogues with updated probabilities of membership. The whole spectroscopic catalogue is available online. P\textsubscript{prev} is the probability assigned in the original article, while P\textsubscript{new} is the probability derived in this work. Columns \#3$-$\#7 correspond to the data-set of B11 (heliocentric velocity, equivalent width of the Mg {\sc i} line, metallicity, P\textsubscript{prev}, P\textsubscript{new}), while \#8$-$\#11 to the one of W09 (heliocentric velocity, pseudo-equivalent width of the Mg-triplet absorption feature,  P\textsubscript{prev}, P\textsubscript{new}). With respect to the data-set of B11, the error in the equivalent width of the integrated fit of the Mg {\sc i} line was updated from $\sigma_{EW}$=2.8/(S/N) to $\sigma_{EW}$=3.6/(S/N) (\citeauthor{battaglia12} \citeyear{battaglia12}), while the metallicity and its error were derived as in B11, from the near-infrared Ca {\sc ii} triplet (CaT) region adopting the calibration of \cite{starkenburg10}. As B11 did not assign proper membership probabilities, just member/nonmember designation, we associated their members and nonmembers with membership probabilities of 1 and 0 respectively.}
		\label{spectroscopic_catalogue}
		\centering
		\begin{tabular}{ r c c c c c c c c c c c}
			\hline\hline
			\noalign{\smallskip}
			\multicolumn{1}{r}{ID} & \multicolumn{2}{c}{Coordinates} & \multicolumn{5}{c}{Data-set from \cite{battaglia11}} & \multicolumn{4}{c}{Data-set from \cite{walker09}}\\
			\hline
			\noalign{\smallskip}
			&   $\alpha_{2000}$ & $\delta_{2000}$ & v\textsubscript{hel} & EW Mg & [Fe/H] & P\textsubscript{prev} & P\textsubscript{new} & v\textsubscript{hel} & $\Sigma$Mg & P\textsubscript{prev} & P\textsubscript{new}\\    
			&  (hh:mm:ss) & (dd:mm:ss) & (km s\textsuperscript{-1}) & (\AA) & (dex)  &  &  & (km s\textsuperscript{-1}) & (\AA)\\
			\noalign{\smallskip}
			\hline
			\noalign{\smallskip}
			1 & 10:11:35.40 & -01:58:13.7 & 214.23$\pm$1.11 & 0.007$\pm$0.151 & $-2.42\substack{+0.14\\-0.15}$ & 1 & 0.999 & 217.4$\pm$1.4 & 0.42$\pm$0.05 & 0.985 & 0.995\\
			2 & 10:12:43.34 & -01:32:05.2 & 56.51$\pm$1.40 & 1.192$\pm$0.114 & $-1.18\substack{+0.08\\-0.08}$ & 0 & 0.000 & 56.3$\pm$0.7 & 0.93$\pm$0.02 & 0.000 & 0.000\\	
			3 & 10:12:32.66 & -02:00:05.2 & 220.25$\pm$2.56 & -0.526$\pm$0.233 & $-1.75\substack{+0.20\\-0.21}$ & 1 & 0.976 & 222.7$\pm$2.5 & 0.80$\pm$0.08 & 0.002 & 0.545\\			
			4 & 10:12:53.46 & -01:15:16.5 & 230.86$\pm$1.87 & 0.184$\pm$0.217 & $-2.86\substack{+0.27\\-0.36}$ & 1 & 0.999 & 233.5$\pm$2.0 & 0.02$\pm$0.07 & 0.738 & 0.999\\	
			5 & 10:13:00.59 & -01:25:47.0 & 111.31$\pm$0.68 & 0.448$\pm$0.057 & $-1.94\substack{+0.04\\-0.05}$ & 0 & 0.000 & 109.7$\pm$0.7 & 0.88$\pm$0.01 & 0.000 & 0.000\\		
			6 & 10:13:01.59 & -01:21:49.9 & -18.28$\pm$2.87 & 0.842$\pm$0.180 & $-1.36\substack{+0.14\\-0.15}$ & 0 & 0.000 & -18.0$\pm$0.7 & 0.99$\pm$0.04 & 0.000 & 0.000\\		
			... & ... & ... & ... & ... & ... & ... & ... & ... & ... & ... & ...\\
			\noalign{\smallskip}
			\hline   
		\end{tabular}
	\end{table*}
	
	\section{Summary and conclusions}
	\label{sec:conclusions}
	
	We present results from CTIO/DECam deep $g-$ and $r-$band photometry of the Sextans dSph out to a radius of $\sim$4 deg, covering approximately 20 deg$^2$ and reaching $\sim$ 2 magnitudes below the oldest main-sequence turn-off. The photometric catalogue of point-like sources is made publicly available.
	
	We updated the structural parameters of the galaxy (Table \ref{structural_parameters}) by fitting different surface density profiles through a Bayesian MCMC sampling of the likelihood evaluated at each star's location. We find overall agreement with the structural analysis of IH95, but with Sextans having a smaller half-light radius than previously reported by IH95 and R16. Likewise, through the Posterior Bayes Factor we established that the best-fitting King profile and corresponding structural parameters from this work are strongly favoured over the best-fitting models by IH95 and R16. In addition, we updated the apparent, absolute magnitude and central surface brightness in V-band: $V=10.73\substack{+0.06\\-0.05}$ mag, $M_V=-8.94\substack{+0.11\\-0.09}$ mag and $\mu_V=27.25\substack{+0.06\\-0.05}$ mag/arcsec$^{-2}$ respectively (already corrected for Galactic extinction), which are compatible with the values from \cite{irwin95} but with considerably reduced errors.
	
	We decontaminated two-dimensional surface density maps of Sextans overall stellar population by making use of an improved version of the \cite{mcmonigal14} matched-filter method. Sextans displays a fairly regular distribution with no significant distortions down to a surface brightness level of $\sim$31.8 mag/arcsec$^{-2}$ in V-band, unlike the surface density map generated by \cite{roderick16}. By studying the 2D distribution of residuals with respect to the fitted surface density axysimmetric models, we have detected several over-dense and under-dense clumps at the 2$\sigma$/3$\sigma$ levels from which we identified an overdensity in the galactic centre that might correspond to the cold substructure/s detected by \cite{kleyna04} and \cite{battaglia11}. The origin of the under-dense clumps is still unclear. Stars forming part of these over/under-dense clumps are therefore of particular interest for future spectroscopic studies of Sextans, in particular those stars belonging to the overdensity in its centre that might correspond to the cold substructure/s. We defer to a future paper an in-depth analysis of the possible presence of sub-structures in Sextans' inner regions.
	
	For the first time, we carry out a quantitative determination of the structural properties and number density profiles of stars in different evolutionary phases in Sextans, i.e. RHB, BHB and BS stars. No significant distortions were found for any of these populations. RHB and BHB stars have clearly distinct spatial distributions, with the RHB stars exhibiting a much smaller 2D half-light radius than the BHB ones (17 vs. 42 arcmin, respectively) and slightly smaller than the overall Sextans stellar population (22 arcmin). This is consistent with the age and metallicity gradient found by \cite{okamoto17} and \cite{battaglia11}, respectively, and puts on a quantitative basis the more qualitative type of findings by \cite{bellazzini01}, \cite{lee03}, \cite{lee09}, \cite{roderick16} and \cite{okamoto17}. With regard to BSs, we confirm that the bright BSs ($g<22.3$) are less spatially extended than the faint ($g>22.3$) ones. The compatibility of the spatial distribution of the bright BSs with the RHB stars, and with the faint BSs with the whole population, appears compatible with the hypothesis that the bulk of Sextans BSs evolved from mass transfer of primordial binaries.
	
	Finally, we use the revised Sextans structural properties from the analysis of our photometric data-set to update the membership probabilities of stars in the spectroscopic catalogues by \cite{battaglia11} and \cite{walker09}, following the decontamination methodology of the latter in an improved form. This catalogue is also made publicly available in order to facilitate subsequent studies of the internal properties of Sextans.
	
	\begin{acknowledgements}
		The authors thank the anonymous referee for a thorough reading of the manuscript and suggestions that improved its clarity. L. Cicuéndez Salazar acknowledges hospitality at the Institute of Astronomy, University of Cambridge, during part of this work and financial support by the University of la Laguna thanks to its international mention grant. He also acknowledges Fundanción la Caixa for the financial support received in the form of a PhD contract. G. Battaglia acknowledges financial support by the Spanish Ministry of Economy and Competitiveness (MINECO) under the Ramón y Cajal Programme (RYC-2012-11537).\\	  
		
		This research has been supported by MINECO under the grant \mbox{AYA2014-56795-P}.\\
		
		This project used data obtained with the Dark Energy Camera (DECam), which was constructed by the Dark Energy Survey (DES) collaboration.
		Funding for the DES Projects has been provided by 
		the U.S. Department of Energy, 
		the U.S. National Science Foundation, 
		the Ministry of Science and Education of Spain, 
		the Science and Technology Facilities Council of the United Kingdom, 
		the Higher Education Funding Council for England, 
		the National Center for Supercomputing Applications at the University of Illinois at Urbana-Champaign, 
		the Kavli Institute of Cosmological Physics at the University of Chicago, 
		the Center for Cosmology and Astro-Particle Physics at the Ohio State University, 
		the Mitchell Institute for Fundamental Physics and Astronomy at Texas A\&M University, 
		Financiadora de Estudos e Projetos, Funda{\c c}{\~a}o Carlos Chagas Filho de Amparo {\`a} Pesquisa do Estado do Rio de Janeiro, 
		Conselho Nacional de Desenvolvimento Cient{\'i}fico e Tecnol{\'o}gico and the Minist{\'e}rio da Ci{\^e}ncia, Tecnologia e Inovac{\~a}o, 
		the Deutsche Forschungsgemeinschaft, 
		and the Collaborating Institutions in the Dark Energy Survey. 
		
		The Collaborating Institutions are 
		Argonne National Laboratory, 
		the University of California at Santa Cruz, 
		the University of Cambridge, 
		Centro de Investigaciones En{\'e}rgeticas, Medioambientales y Tecnol{\'o}gicas-Madrid, 
		the University of Chicago, 
		University College London, 
		the DES-Brazil Consortium, 
		the University of Edinburgh, 
		the Eidgen{\"o}ssische Technische Hoch\-schule (ETH) Z{\"u}rich, 
		Fermi National Accelerator Laboratory, 
		the University of Illinois at Urbana-Champaign, 
		the Institut de Ci{\`e}ncies de l'Espai (IEEC/CSIC), 
		the Institut de F{\'i}sica d'Altes Energies, 
		Lawrence Berkeley National Laboratory, 
		the Ludwig-Maximilians Universit{\"a}t M{\"u}nchen and the associated Excellence Cluster Universe, 
		the University of Michigan, 
		{the} National Optical Astronomy Observatory, 
		the University of Nottingham, 
		the Ohio State University, 
		the University of Pennsylvania, 
		the University of Portsmouth, 
		SLAC National Accelerator Laboratory, 
		Stanford University, 
		the University of Sussex, 
		and Texas A\&M University.\\
		
		Based on observations at Cerro Tololo Inter-American Observatory, National Optical Astronomy Observatory (NOAO Prop. ID: 2015A/1013; PI: B. McMonigal), which is operated by the Association of Universities for Research in Astronomy (AURA) under a coope rative agreement with the National Science Foundation.\\
		
		This work has made use of the IAC-STAR Synthetic CMD computation code. IAC-STAR is suported and maintained by the computer division of the Instituto de Astrofísica de Canarias.		
	\end{acknowledgements}
	
	\bibliographystyle{aa}
	
	\bibliography{biblio}
	
	\section*{Appendix A: On the outer regions of DECam pointings}
	
	We attempted to overcome the morphological misclassification of point-like vs extended objects in the out-of-focus regions of the DECam pointings in various ways, testing the outcome of the possible solutions by looking at the resulting surface density maps: our ``figure-of-merit'' was the lack of ring-like over-densities of point-like objects in the external regions of the pointings, above the $\sigma$ detection limits used for the scientific analysis of the density maps (see Sect.\,\ref{sec:maps}).
	
\begin{figure}
		\centering
		\includegraphics[trim={7.75cm 0.5cm 5.75cm 1.25cm},clip,width=0.49\hsize]{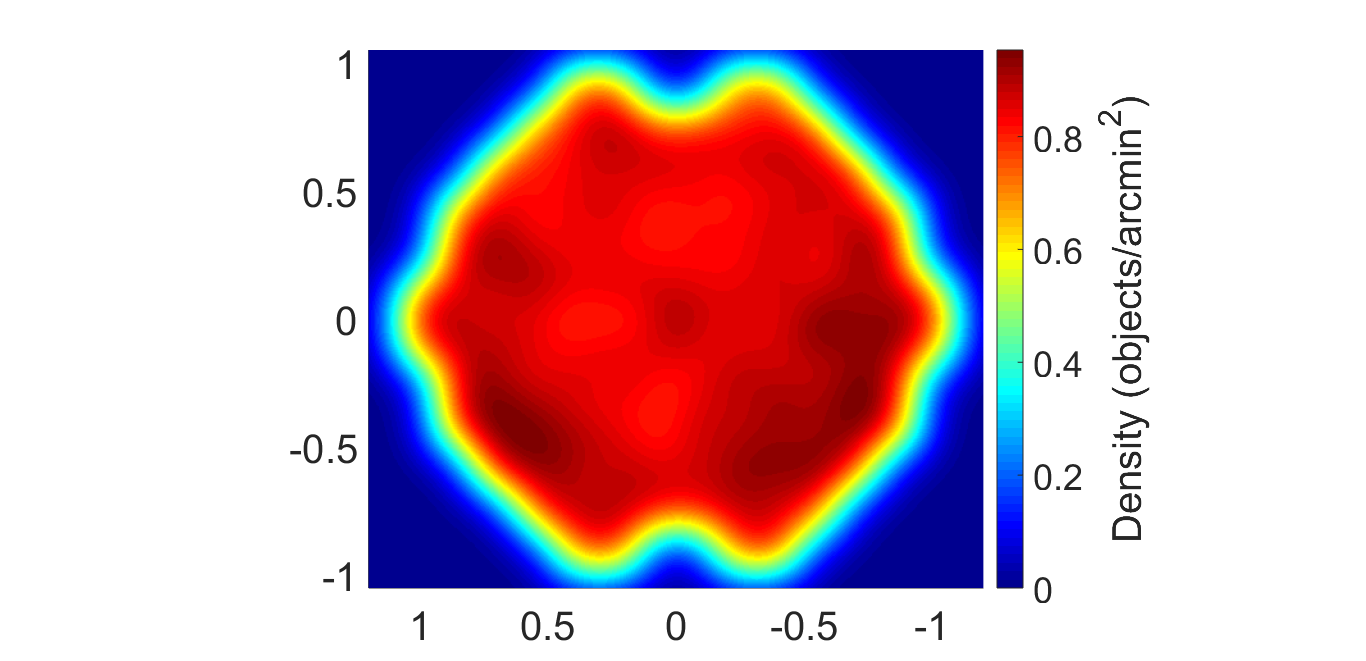}
		\includegraphics[trim={7.75cm 0.5cm 5.75cm 1.25cm},clip,width=0.49\hsize]{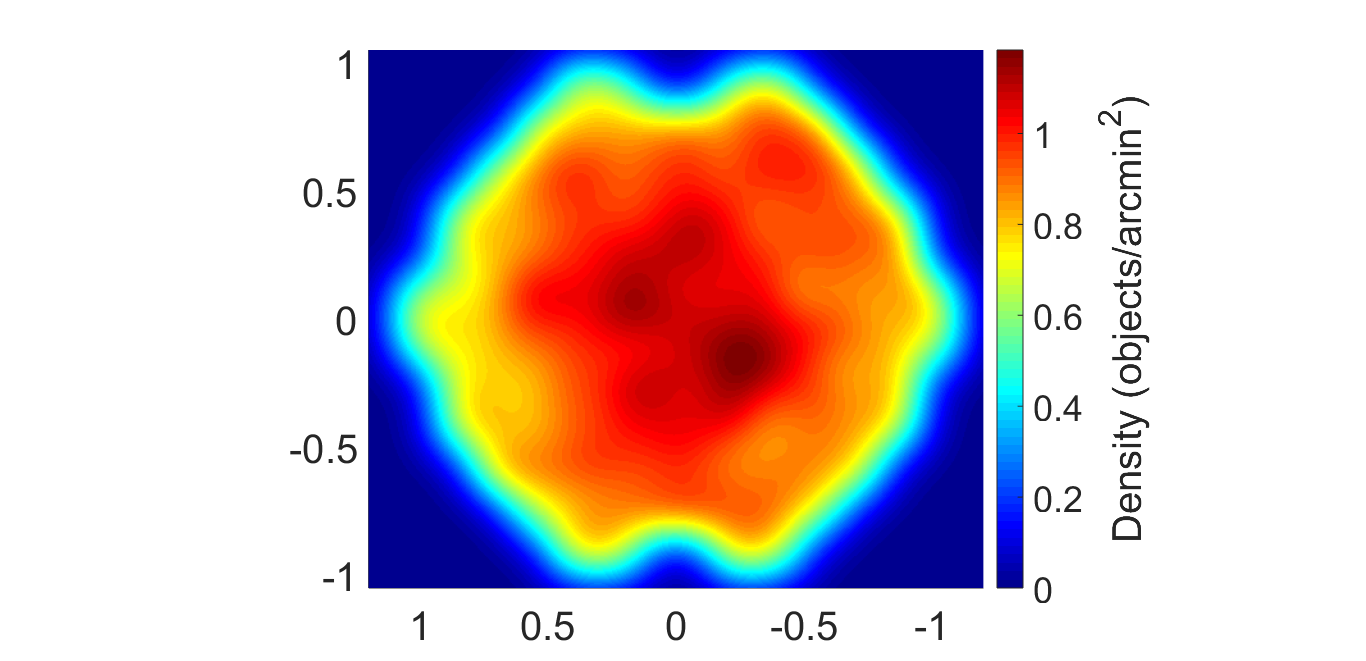}	\includegraphics[trim={2.25cm 0cm 2.75cm 0.5cm},clip,width=\hsize]{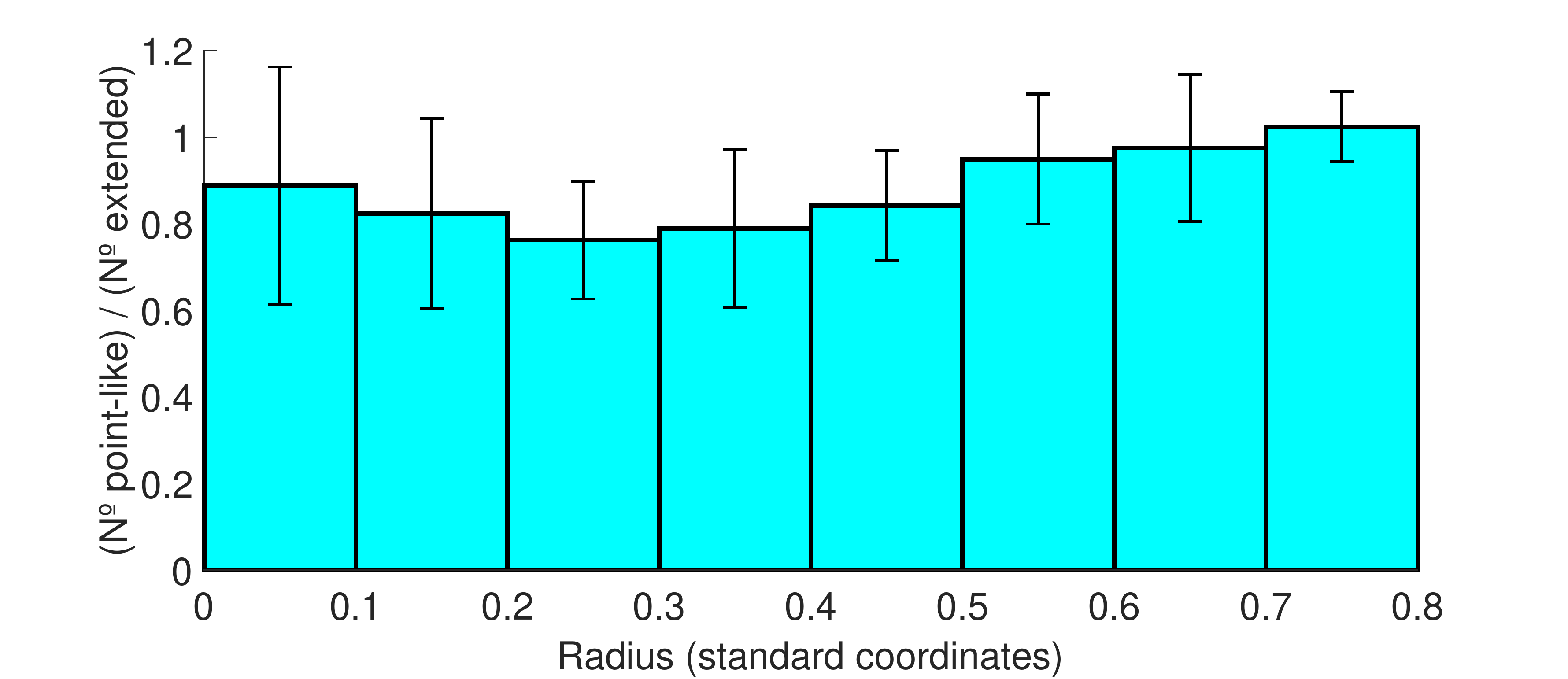}
		\caption{Smoothed map of the mean density pattern for point-like (top-left) and extended (top-right) objects with $1.1<g-r<1.6$ in standard coordinates, discarding $3\sigma$ outliers, with $\sigma$ being the standard deviation between the different pointings. The bottom panel plots the ratio of point-like/extended objects, showing the spatial gradient in the morphological misclassification when moving away from the centre of the pointings. The error-bars indicate the standard deviation of the ratios derived from the different measurements in all the pointings.}
		\label{masks}
	\end{figure}

	We first tried to quantify the morphological misclassification by creating a density map of point sources expected to have a fairly constant spatial distribution over the probed spatial regions i.e. Milky Way foreground stars, which we could then in principle use as a correction factor. To this end, we selected objects in the colour range $1.1<g-r<1.6$, in order to avoid the region of the CMD occupied by Sextans stars (see the top panels of Fig.\,\ref{masks} for the mean density of point-like and extended objects in this colour range). Since the details of the features vary from pointing to pointing, the smoothed density of these objects was derived across the full mosaic area to correct for all artificial inhomogeneities at once. One issue with this approach was that the morphological misclassification depends on magnitude, and the luminosity function both of point-like and extended sources in the colour range  $1.1<g-r<1.6$ is not representative of the luminosity function across the whole colour range of the data (for example in the region occupied by Sextans stars).
	
	\begin{figure*}[!h]
		\centering
		\includegraphics[trim={1cm 3.5cm 0.5cm 5.5cm},clip,width=\hsize]{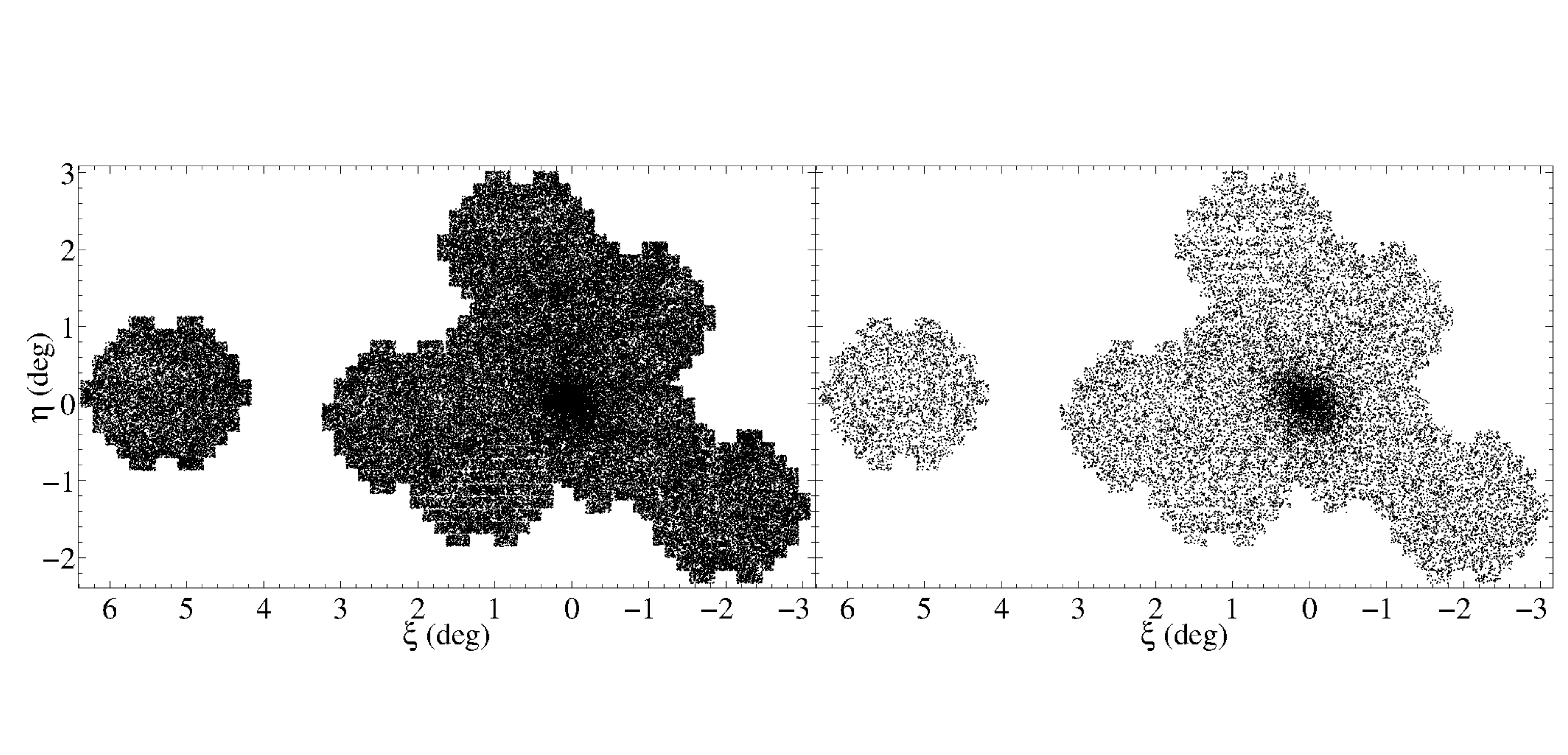}
		\caption{Locations of the objects from our photometric catalogue.
			Left: all sources; right: point-like sources brighter than $(g,r)=(23.0,23.0)$.}
		\label{flags_maps}
	\end{figure*}       
	
	Another attempt was to use the inverse density map of extended objects as a correction factor to the map of point-like sources, since the overdensities of morphologically misclassified point-like sources correspond to underdensities of extended objects. Nonetheless clusters of galaxies and differences in depth between pointings introduce much larger inhomogeneities in the map of extended objects than those created by the star/galaxy misclassification, so applying this mask created even more cosmetic features.
	
	The most satisfactory solution was to use only point-like sources brighter than $(g,r)=(23.0,23.0)$, approximately the region where the locus of unresolved galaxies starts appearing on the CMD. As it can be seen in Fig.\,\ref{flags_maps}, artificial overdensities and clusters of galaxies are effectively removed, though at the cost of reduced statistics. Although the sample under consideration has a brighter magnitude limit than the catalogue used by R16, comparing Fig.\,\ref{fig_fit_king} with their Fig.\,8 one can see that the central densities of both catalogues are very similar ($\sim$4 stars/arcmin$^2$). This implies that, in spite of their deeper magnitude cut, the reached number density of stars from the galaxy (and therefore the statistics) is similar in both data-sets, possibly due to distinct selection criteria compensating the different magnitude cuts.
	
	When attempting to correct for artificial features, the aforementioned masks were applied in the form of a ``flat-field'', i.e. dividing the density maps of point-like sources by the normalized masks. In R16 the mask for the whole mosaic created from stars outside their CMD window (likely contaminants) was applied to the normalized density map of the stars falling inside it (likely Sextans' stars) in the form of a flat-field too; however, it was then also subtracted from the resulting ``flat-fielded'' map to emphasize low-density features of Sextans' stars. In our opinion this last step may strongly affect the real shape of the dSph by underestimating the galactic density in the most contaminated regions, regardless of whether their galaxy/contamination density ratio is high or low. This would create holes in the contour map of the galaxy, mimicking the appearance of over-dense distortions surrounding the overdensities of contamination, where instead the galactic density can remain constant. We think this may contribute to the boxy appearance of Sextans in R16, aligned with its major and minor axes; this would be caused by the higher detected contamination density in the overlapping regions of the pointings placed along the major and minor axes of Sextans.	
	
	\section*{Appendix B: MCMC Hammer applied to mock galaxies}
	
	We test the performance of the MCMC Hammer method used in Sect.\,\ref{sec:structural}	by applying it to mock data-sets of point-sources, aimed at probing different regimes of number statistics and spatial coverage of the data-set.
	
	First, we generate catalogues of point-sources distributed according to a given surface density profile $f(r)$ following the method outlined in \textit{Numerical Recipes} (\citeauthor{press92} \citeyear{press92}) for the generation of random numbers with a desired distribution. Once we have calculated the corresponding major axis radii $r_i$ for an exponential profile of scale radius $r_e$, we generate the angular coordinates following a uniform distribution between 0 and 2$\pi$, transform the angular coordinates into cartesian ones, adjust the ellipticity ($e$) by multiplying the vertical coordinates by the factor $(1-e)$ and finally rotate all the data points by the position angle $\theta$. This produces a mock galaxy with an exponential surface density profile centered on the origin. In order to simulate realistic observational conditions, we also added a constant contamination density by generating random stars with a uniform spatial distribution.
	
	The exponential radius and position angle are fixed to $r_e=$10 arcmin and $\theta=50^{\circ}$, respectively. The parameters we vary are the number of mock stars within the spatial coverage of the mock data-set, $N$, the ellipticity of the galaxy, $e$, and the coverage. We generate 18 mock galaxies corresponding to a grid of 3x2x3 in the aforementioned parameters ($N=$ 200, 1000, 5000; $e=$ 0.1, 0.3; for the coverage we used squares of sides equal to 2, 5 and 10x$r_e$). We set the number of contaminants to
	20\% of the total number $N$ of stars, which is fixed inside the different coverages. This results into different contamination densities and scale factors $k$ (equal to the central densities in the case of the exponential profile) between the different coverages with same $N$. As when fitting the data-set of the Sextans dSph, in the MCMC Hammer we defined 80 walkers, each of them doing approximately 10$^4$ steps.
	
	Table \ref{mocks} summarizes the best-fitting parameters we obtain. As expected, there is a clear effect of the ellipticity on the determination of position angles, especially at low coverages and number of data points. Likewise, the effect of the area is important since low coverages produce difficulties in fitting the contamination density when there are no areas with low density of galaxy population. Even if the number of analysed stars is high, as with $N$ = 5000, there are still high dispersions in this parameter at 2$r_e$ coverage. The scale length, the position angle and the ellipticity are also quite affected by the size of the analysed area, having higher error bars at low coverages.	In the case of the data-set of Sextans we had an ellipticity close to 0.3, a total number of stars from the galaxy near 5000 and a coverage larger than 10$r_e$. These numbers make the data-set of Sextans similar to the mock one with the structural parameters most constrained by the MCMC Hammer.
	
	The large error bars in some parameters can make the Bayesian posterior distribution non-gaussian when there are prior restrictions on the parameter. This happens with the central surface density and contamination density, which are set to be greater than zero in order to have physical meaning; and with the ellipticity, that is defined between 0 and 1. However, what really matters for this work is how reliable the inferred error-bars on the parameters estimates are. {\it In this respect Table \ref{mocks} shows that, if the models are a good representation of the data, we can rely on the error bars, since the input parameters for all the mock galaxies are always recovered within the 2$\sigma$ confidence level and in most cases within the 1$\sigma$ level, even for low number statistics and restricted spatial coverages; the trade-off being much larger error-bars, obviously.}
	
	\begin{table}
		\caption{Structural parameters (median values of the marginalized posterior distributions) obtained with the MCMC Hammer for mock galaxies with number of analysed stars $N$ = 200, 1000 and 5000 (top, middle and bottom sequences of rows respectively). From left to right: input ellipticity ($e_i=1-b/a$), coverage (in $r_e$ units), input contamination density ($\sigma_{c,\,i}$) and fitted parameters: scale factor ($k$, equal to the central density in the case of the exponential profile), exponential radius ($r_e$), galactic centre ($\alpha_0$, $\delta_0$), position angle ($\theta$), ellipticity ($e$) and contamination density ($\sigma_{c}$).}           
		\label{mocks}
		\tiny
		\centering
		\begin{tabular}{c c c c c c c}        
			\hline\hline                 
			\noalign{\smallskip}
			$e_i$ & Side & $\sigma_{c,\,i}$  & $r_e$  & $\theta$ & $e$ & $\sigma_c$\\
			& ($r_e$)  & (\#/arcmin$^2$)  & (') & (º) &   & (\#/arcmin$^2$)\\
			\noalign{\smallskip}
			\hline                        
			\noalign{\smallskip}
			0.1 & 2 &  0.100
			& 29.1$\substack{+89.7\\-17.7}$  &  63$\substack{+31\\-33}$ & 0.45$\substack{+0.31\\-0.29}$ & 0.209$\substack{+0.166\\-0.142}$ \\
			\noalign{\smallskip}
			0.1 & 5 & 0.016
			& 8.7$\substack{+1.9\\-1.6}$  & 48$\substack{+32\\-31}$ & 0.11$\substack{+0.10\\-0.08}$ & 0.022$\substack{+0.011\\-0.011}$ \\
			\noalign{\smallskip}
			0.1 & 10 & 0.004
			& 9.8$\substack{+1.6\\-1.4}$  &  65$\substack{+32\\-38}$ & 0.10$\substack{+0.11\\-0.07}$ & 0.007$\substack{+0.002\\-0.002}$ \\
			\noalign{\smallskip}
			0.3 & 2 & 0.100
			& 18.7$\substack{+61.5\\-11.9}$  &  53$\substack{+32\\-40}$ & 0.39$\substack{+0.30\\-0.25}$ & 0.258$\substack{+0.169\\-0.174}$ \\
			\noalign{\smallskip}
			0.3 & 5 & 0.016
			& 7.5$\substack{+1.9\\-1.5}$  &  35$\substack{+15\\-16}$ & 0.26$\substack{+0.13\\-0.13}$ & 0.035$\substack{+0.010\\-0.010}$ \\
			\noalign{\smallskip}
			0.3 & 10 & 0.004
			& 10.6$\substack{+1.1\\-1.0}$  &  51$\substack{+12\\-13}$ & 0.21$\substack{+0.09\\-0.10}$ & 0.003$\substack{+0.001\\-0.001}$ \\
			\noalign{\smallskip}
			\hline   
			\noalign{\smallskip}
			0.1 & 2 & 0.50
			& 8.9$\substack{+2.7\\-2.7}$  & 73$\substack{+18\\-27}$ & 0.16$\substack{+0.14\\-0.10}$ & 0.74$\substack{+0.60\\-0.50}$\\
			\noalign{\smallskip}
			0.1 & 5 & 0.08
			& 10.0$\substack{+1.5\\-1.4}$  & 40$\substack{+15\\-13}$ & 0.14$\substack{+0.06\\-0.07}$ & 0.09$\substack{+0.04\\-0.04}$\\
			\noalign{\smallskip}		
			0.1 & 10 & 0.02
			& 9.4$\substack{+0.5\\-0.5}$ & 43$\substack{+17\\-18}$ & 0.08$\substack{+0.05\\-0.04}$ & 0.02$\substack{+0.01\\-0.01}$\\
			\noalign{\smallskip}
			0.3 & 2 & 0.50
			& 9.4$\substack{+2.3\\-2.3}$  & 47$\substack{+21\\-21}$ & 0.17$\substack{+0.10\\-0.10}$ & 0.58$\substack{+0.53\\-0.40}$\\
			\noalign{\smallskip}
			0.3 & 5 & 0.08
			& 10.2$\substack{+0.9\\-0.8}$  & 46$\substack{+4\\-4}$ & 0.38$\substack{+0.04\\-0.04}$ & 0.07$\substack{+0.02\\-0.02}$\\
			\noalign{\smallskip}
			0.3 & 10 & 0.02
			& 10.9$\substack{+0.6\\-0.6}$  & 51$\substack{+3\\-3}$ & 0.36$\substack{+0.04\\-0.04}$ & 0.02$\substack{+0.01\\-0.01}$\\
			\noalign{\smallskip}
			\hline
			\noalign{\smallskip}
			0.1 & 2 & 2.50
			& 11.3$\substack{+1.4\\-1.6}$  & 52$\substack{+16\\-19}$ & 0.10$\substack{+0.06\\-0.05}$ & 1.41$\substack{+1.65\\-1.03}$\\
			\noalign{\smallskip}
			0.1 & 5 & 0.40
			& 10.3$\substack{+0.7\\-0.6}$  & 47$\substack{+21\\-21}$ & 0.04$\substack{+0.03\\-0.03}$ & 0.32$\substack{+0.10\\-0.10}$\\
			\noalign{\smallskip}
			0.1 & 10 & 0.10
			& 10.0$\substack{+0.2\\-0.2}$  & 54$\substack{+7\\-7}$ & 0.09$\substack{+0.02\\-0.02}$ & 0.10$\substack{+0.01\\-0.01}$\\
			\noalign{\smallskip}
			0.3 & 2 & 2.50
			& 10.6$\substack{+1.8\\-1.8}$  & 48$\substack{+6\\-6}$ & 0.30$\substack{+0.04\\-0.04}$ & 2.52$\substack{+1.69\\-1.65}$\\
			\noalign{\smallskip}
			0.3 & 5 & 0.40
			& 10.3$\substack{+0.5\\-0.5}$  & 47$\substack{+3\\-3}$ & 0.29$\substack{+0.02\\-0.02}$ & 0.40$\substack{+0.06\\-0.06}$\\
			\noalign{\smallskip}
			0.3 & 10 & 0.10
			& 10.1$\substack{+0.2\\-0.2}$  & 49$\substack{+2\\-2}$ & 0.31$\substack{+0.02\\-0.02}$ & 0.10$\substack{+0.01\\-0.01}$\\
			\noalign{\smallskip}
			\hline
		\end{tabular}
	\end{table}

	\section*{Appendix C: Goodness-of-fit indicators}
	
	The $\chi^2_{red}$ was calculated from the observed surface number density profile derived from elliptical, concentric annuli and the best-fitting profiles, by assuming that star counts follow Poisson distributions with means equal to the star counts from the best-fit model.
	
	Although the $\chi^2_{red}$ gives a rough description of how good a fitting is, it is not always a reliable discriminant to probe which model better fits the data: it is required to be in a regime in which the Poisson distribution can be approximated by a normal distribution and the value is quite dependent on the choice of the bin size. For the same reasons, structural parameters derived via conventional $\chi^2_{red}$ fitting are not as reliable as when they are derived following more appropriated techniques. As one can check in Table \ref{structural_parameters}, according to their $\chi^2_{red}$ all the functional forms of the profiles fitted in this work perform approximately equally well, even when accounting for the profiles fitted by IH95. As we will see later, the order given by $\chi^2_{red}$ does not coincide with the ones given by more reliable indicators. In this respect the ratio of maximum likelihoods tell us how many times the data is more likely to follow one profile than the other, without making assumptions or loosing information due to the spatial binning.
	
	Nonetheless, neither the $\chi^2$ or the classical likelihood ratio take into account the probability of having structural parameters different from the most probable ones. Here is where Bayesian statistics makes the difference. Bayes' theorem states:
	\begin{equation}
	P(M|D)=\frac{P(D|M) \, P(M)}{P(D)}\,,
	\end{equation}
	where the events $M$ and $D$ are, in the case that concerns us, the model being evaluated (i.e. the surface density profile) and our data-set respectively. $P(M|D)$, known as the posterior, is the probability of the analysed profile to reproduce our data-set. $P(D|M)$, known as the likelihood, is the probability of our data-set to reproduce the profile. $P(M)$, known as the prior, is the probability of the profile regardless of our specific data-set. Finally, the normalizing constant $P(D)$, known as the marginal likelihood, is the global probability of our data-set summing the probabilities of all the considered profiles. In the same way, we can use Bayes' theorem to calculate the posteriors of the structural parameters from a given profile, assuming that the considered profile is the true one:
	\begin{equation}
	P(\theta|M,D)=\frac{P(D|M,\theta) \, P(\theta|M)}{P(D|M)}\,,
	\label{eq:post_para}
	\end{equation}
	with $\theta$ being the structural parameters and the marginal likelihood being: 
	\begin{equation}
	P(D|M)=\int_\theta \! P(D|M,\theta) \, P(\theta|M) \, \mathrm{d}\theta\,,
	\label{eq:mar_like}
	\end{equation}
	integrated over all the structural parameters' domain. 
	
	Integrating Eq.\,\ref{eq:post_para} along its different dimensions $\theta_i$ we can then obtain the variety of projections previously shown in Fig.\,\ref{fig_king_posteriors}.
	
	Through Eq.\,\ref{eq:mar_like}, i.e. marginalizing the likelihood over all its parameters, one considers all the possible shapes that each profile can take, depending on the probability distribution functions of the model parameters. This can potentially result in the fact that a model much more likely than others when considering the most probable values of its parameters, is less likely when considering the full distribution of possible values. This is the Occam's razor intrinsic to Bayesian statistics, which automatically penalizes overfitted profiles due to an excess of variables.
	
	The ratio between the posteriors of two different models tell us how many times one model is more likely to follow the data than the other. If we do not have a prior idea of which model is more likely, both priors in the expression of the posteriors vanish when dividing, which results in the ratio of the marginal likelihoods of both models. This ratio between the marginal likelihoods of two different models is called the Bayes Factor:
	\begin{equation}
	\frac{P(M_1|D)}{P(M_2|D)}=\frac{P(D|M_1)}{P(D|M_2)}=K_{12}  \quad \mbox{if} \; P(M_1)=P(M_2)
	\end{equation}
	
	However, it is not straightforward to evaluate these marginal likelihoods over the high dimensional spaces of the fitted profiles. In computational terms it is not feasible to calculate the numerical integral of Eq.\,\ref{eq:mar_like} over their corresponding N-dimensional grids; they need to be evaluated through Monte Carlo integration (\citeauthor{robert04} \citeyear{robert04}). Here is where the MCMC becomes important: given two functions, $f(\theta)$ and $g(\theta)$, with the last one being normalized, the integral $\int_\theta f(\theta) \, g(\theta) \, \mathrm{d}\theta$ is equal to the mean value of $f(\theta)$ evaluated at the points sampled with the MCMC from $g(\theta)$. Thus, in principle, we could obtain the Bayes' factors between the different profiles by calculating Eq.\,\ref{eq:mar_like} for each profile through the MCMC Hammer and the expressions of the likelihood $P(D|M,\theta)=f(\theta)$ and prior $P(\theta|M)=g(\theta)$.
	
	However, if there is any improper function in the definition of some prior there is no possibility of sampling it with a MCMC. As this is our case, we decided to use the alternative of the Posterior Bayes Factor (PBF), defined by \cite{aitkin91}, which has the equivalent definition of the Bayes factor but replacing the prior $P(\theta|M)$ in Eq.\,\ref{eq:mar_like} by the posterior $P(\theta|M,D)$. As we run the MCMC Hammer to calculate the posterior distribution of the model parameters, the PBFs are just the ratios between the means of the likelihoods evaluated at the points sampled with the MCMC for each profile (Fig.\,\ref{fig_king_posteriors}). The PBF will be the most reliable discriminant when deciding how good the different profiles fit the data, as it does not require any assumptions, spatial binning or unique possible values for the structural parameters.
	
	\cite{aitkin91} associated the values of the PBF greater than 20, 100 or 1000 with strong, very strong and overwhelming sample evidence against the less probable model. However, we prefer to use the later widely cited logarithmic scale defined by \cite{kass95} (Table \ref{kass_class}) to list the PBFs and classical likelihood ratios in Table \ref{structural_parameters}.
	\begin{table}
		\caption[.]{Evidence classification of the Bayes factors according to \cite{kass95}}
		\label{kass_class}
		\centering                          
		\begin{tabular}{c c c}       
			\hline\hline                 
			\noalign{\smallskip}
			$2\ln{K_{12}}$ & $K_{12}$ & Evidence against $M_2$\\    
			\noalign{\smallskip}
			\hline                        
			\noalign{\smallskip}
			0 to 2 & 1 to 3 & Not worth more than a bare mention \\
			\noalign{\smallskip}
			2 to 6 & 3 to 20 & Positive \\
			\noalign{\smallskip}
			6 to 10 & 20 to 150 & Strong \\
			\noalign{\smallskip}
			>10 & >150 & Very strong \\
			\noalign{\smallskip}
			\hline                                   
		\end{tabular}
	\end{table}
	
	This logarithmic scale allows us to infer the PBFs between pairs of profiles that were not matched in Table \ref{structural_parameters}, by adding the values corresponding to the pairs that are in between. For example, while the King profile has $2\ln{\textnormal{(PBF)}}=1.6$ against the exponential one, it has $1.6+0.8=2.4$ against the Sérsic profile and $1.6+0.8+1.8=4.2$ against the Plummer. Thus, whereas the evidence is not worth more than a bare mention respect to the exponential profile, there is a positive evidence against the Sérsic and Plummer profiles. Likewise there is a positive evidence in favour of the exponential profile against the Plummer one. Thereby one can check that the order following the goodness-of-fit of the different profiles according to the PBFs is different than that given by the $\chi^2_{red}$ (or even by the likelihood ratio, see e.g. Table \ref{structural_parameters_deep} in Appendix C). Regarding the rest of the matches between the profiles fitted by us their evidence is not worth more than a bare mention, while all the PBFs for the profiles fitted by IH95 and R16 present a very strong (overwhelming in the scale of \cite{aitkin91}) evidence in favour of any of the profiles fitted in this work, with both the $\chi^2_{red}$ and classical likelihood ratio strongly disfavouring their structural parameters too.
	
	\section*{Appendix D: Structural parameters from S/N=5 catalogue}
	
	Here we show the results of applying the analysis from Sect.\,\ref{sec:structural} on the photometric catalogue cut at S/N=5 of the shallowest pointing, i.e. $(g,r)=(24.9,24.9)$, with $\sim$ 440,000 objects instead of the $\sim$ previous 75,000 but affected by spatially variable star/galaxy classification and completeness. Since the parameters giving the direction of the spatial gradient in the density of contaminants turn out to be sensitive to the differences in depth between pointings and possibly the presence of the artificial over-densities in the out-of-focus regions, we adopt the parameters {\it a} and {\it b} from our baseline ``bright'' catalogue.
	
	Also in this case the King model provides the best description of the data. However, as summarized in Table \ref{structural_parameters_deep}, the ellipticity and the values of the scale-lenghts (and the corresponding half-light radii) here derived return a rounder and more compact structure than when using the shallower catalogue. For these parameters, the differences with the values in Tab.\,\ref{structural_parameters} appear to be statistically significant, as they are beyond 4$\sigma$; however these error-bars should be taken with caution: these are ``formal'' error-bars, likely underestimated due to the implicit assumption that the model is a good representation of the data-set, which in the case of this deeper catalogue affected by spatially variable star/galaxy classification and completeness is certainly broken to some extent. 
	
	Interestingly though, the King tidal radius drops to 62 arcmin ($\sim$ 1.6 kpc), consistent within 3$\sigma$ from the determination from the shallower catalogue (120$_{-20}^{+20}$ arcmin). This change in the tidal radius would sweep away one of the most remarkable features of this dSph.
	
	It is hard to pinpoint to what extent the artifacts from the out-of-focus regions and small differences in depth between the pointings are affecting the values here determined. Hence we still recommend using the values in Table \ref{structural_parameters} as they come from a uniform mosaic at the $\sim$ 100\% completeness level.
	
\begin{figure}
		\centering
		\includegraphics[trim={2.25cm 0cm 8.75cm 0cm},clip,width=\hsize]{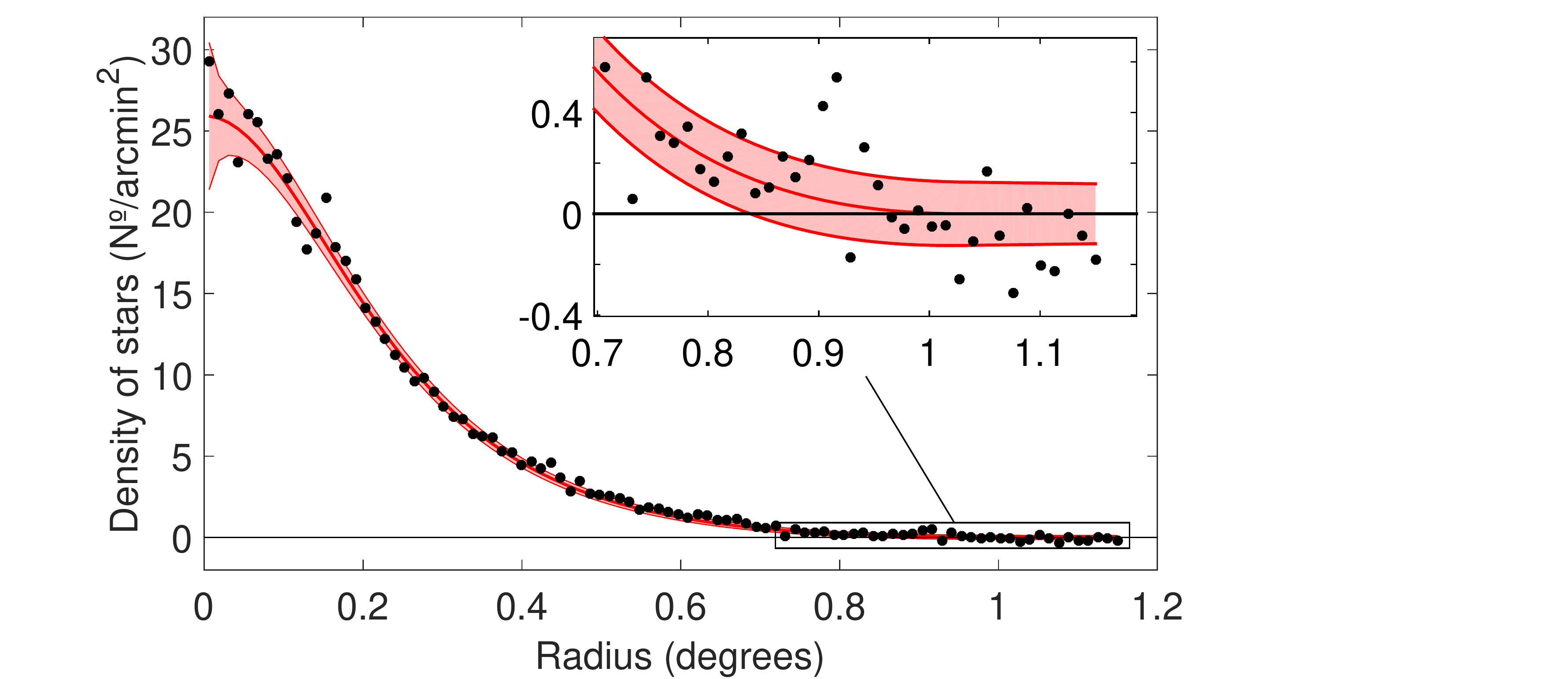} 
		\caption{Contamination subtracted surface number density profile of Sextans stars as a function of the major axis radius (with the external parts zoomed in) from the photometric catalogue cut at S/N=5 from the shallowest DECam pointing, overlaid onto the $1\sigma$ confidence interval (red band) of the best-fitting King profile obtained with the MCMC Hammer. The $1\sigma$ confidence interval is computed from the best-fitting model assuming Poisson variances in each elliptical annulus. }
		\label{fig_fit_king_deep}
	\end{figure}

	Nonetheless, the values of the structural parameters common to the different fitted profiles remain acceptably compatible between them and Fig.\,\ref{fig_fit_king_deep} shows a good agreement between the best-fitting King model and the observed density profile at this depth. This smaller value for the tidal radius agrees quite well with the limits of the decontaminated surface density map (top panel of Fig.\,\ref{mfm_map}) and the decontaminated CDF of the whole population of Sextans, which reaches 1 at a major axis radius of 62 arcmin, not increasing for larger distances; these could be seen as consistency checks since the tidal radius, the density maps and the CDFs are inferred by independent methods. We also note that this tidal radius would also agree with the extension of the spectroscopic samples (left panel of Fig.\,\ref{prob_spec}), not detecting any star with high probability of membership beyond it.
	
	\begin{table*}
		\caption[.]{Sextans structural parameters (median values of the marginalized posterior distributions) derived with the MCMC Hammer applied to the photometric catalogue cut at S/N=5 from the shallowest DECam pointing, plus $\chi^2_{red}$, classical likelihood ratios and Posterior Bayes Factors of the different surface density profiles. Classical likelihood ratios and PFBs are colour-coded as in Fig.\,\ref{structural_parameters}.}           
		\label{structural_parameters_deep}      
		\centering                          
		\begin{tabular}{c c c c c c c}        
			\hline\hline                 
			\noalign{\smallskip}
			Parameter & Exponential & Sérsic & Plummer & King & IH95 & R16\\    
			\noalign{\smallskip}
			\hline                        
			\noalign{\smallskip}
			$\alpha_{2000}$ (º) & $153.256\substack{+0.002\\-0.003}$ & $153.255\substack{+0.003\\-0.003}$ & $153.256\substack{+0.003\\-0.003}$ & $153.256\substack{+0.003\\-0.003}$ & $153.2625\substack{+0.0005\\-0.0005}$ & $153.277\substack{+0.003\\-0.003}$ \\      
			\noalign{\smallskip}
			$\delta_{2000}$ (º) & $-1.623\substack{+0.002\\-0.003}$ & $-1.623\substack{+0.002\\-0.002}$    & $-1.623\substack{+0.003\\-0.002}$ & $-1.623\substack{+0.003\\-0.002}$ & $-1.6147\substack{+0.0003\\-0.0003}$ & $-1.617\substack{+0.008\\-0.008}$ \\
			\noalign{\smallskip}
			Ellipticity & $0.11\substack{+0.02\\-0.02}$ & $0.13\substack{+0.02\\-0.02}$ & $0.13\substack{+0.02\\-0.02}$ & $0.15\substack{+0.02\\-0.02}$ & $0.35\substack{+0.05\\-0.05}$ & $0.29\substack{+0.03\\-0.03}$ \\
			\noalign{\smallskip}
			Position angle (º) & $50\substack{+3\\-3}$ & $53\substack{+3\\-3}$ & $50\substack{+3\\-3}$ & $53\substack{+3\\-3}$ & $56\substack{+5\\-5}$ & $58\substack{+6\\-6}$\\
			\noalign{\smallskip}
			Sérsic index n & - & $0.71\substack{+0.02\\-0.02}$ & - & - & - & - \\
			\noalign{\smallskip}
			Sérsic factor b(n) & - & $0.9\substack{+0.5\\-0.4}$ & - & - & - & - \\		
			\noalign{\smallskip}
			Exponential $r_{e}$ ($'$) & $10.6\substack{+0.2\\-0.2}$ & - & - & - &  $15.5\substack{+0.1\\-0.1}$ & - \\
			\noalign{\smallskip}
			Plummer $r_{p}$ ($'$) & - & - & $19.3\substack{+0.3\\-0.3}$ & - & - & $35.7\substack{+0.7\\-0.7}$\\
			\noalign{\smallskip}
			Sérsic $r_{s}$ & - & $15\substack{+5\\-5}$ & - & - & - & - \\
			\noalign{\smallskip}
			King $r_{c}$ ($'$) & - & - & - & $16.7\substack{+0.5\\-0.5}$ & $17\substack{+2\\-2}$ & $27\substack{+2\\-2}$ \\
			\noalign{\smallskip}
			King $r_{t}$ ($'$) & - & - & - & $62\substack{+2\\-2}$ & $160\substack{+50\\-50}$ & $83\substack{+8\\-8}$ \\
			\noalign{\smallskip}
			2D Half-light $r_{h}$ ($'$) & $17.8\substack{+0.2\\-0.2}$ & $15\substack{+5\\-5}$ & $19.3\substack{+0.3\\-0.3}$ & $17\substack{+1\\-1}$ & $26\substack{+0.2\\-0.2}$ (Exp.) &  $35.7\substack{+0.7\\-0.7}$ (Plummer) \\
			& & & & & $28\substack{+5\\-5}$ (King) & $24\substack{+2\\-2}$ (King) \\
			\noalign{\smallskip}
			$\chi^2_{red}$ & 1.67& 1.37 & 1.53 & 1.31 & 3.95 $\left(\frac{\textnormal{Exp. IH95}}{\textnormal{Exp.}}\right)$ & 15.55 $\left(\frac{\textnormal{Plummer R16}}{\textnormal{Exp.}}\right)$ \\ & & & & & 4.59 $\left(\frac{\textnormal{King IH95}}{\textnormal{Exp.}}\right)$ & 7.48 $\left(\frac{\textnormal{King R16}}{\textnormal{Exp.}}\right)$ \\
			\noalign{\smallskip}
			$2\ln{\left(\frac{\textnormal{likelihood}_{1}}{\textnormal{likelihood}_{2}}\right)}$ & - & \textcolor{red}{102.2} $\left(\frac{\textnormal{Sérsic}}{\textnormal{King}}\right)$ & \textcolor{red}{10.5} $\left(\frac{\textnormal{Plummer}}{\textnormal{Exp.}}\right)$ & \textcolor{red}{94.2} $\left(\frac{\textnormal{King}}{\textnormal{Plummer}}\right)$ & \textcolor{red}{-1225.1} $\left(\frac{\textnormal{Exp. IH95}}{\textnormal{Exp.}}\right)$ & \textcolor{red}{-3452.1} $\left(\frac{\textnormal{Plummer R16}}{\textnormal{Exp.}}\right)$ \\ & & & & & \textcolor{red}{-1355.7} $\left(\frac{\textnormal{King IH95}}{\textnormal{Exp.}}\right)$ & \textcolor{red}{-1468.8} $\left(\frac{\textnormal{King R16}}{\textnormal{Exp.}}\right)$ \\
			\noalign{\smallskip}
			$2\ln{\textnormal{(PBF)}}$ & - & \textcolor{red}{195.4} $\left(\frac{\textnormal{Sérsic}}{\textnormal{Plummer}}\right)$ & \textcolor{red}{10.2} $\left(\frac{\textnormal{Plummer}}{\textnormal{Exp.}}\right)$ & \textcolor{red}{22.7} $\left(\frac{\textnormal{King}}{\textnormal{Sérsic}}\right)$ & \textcolor{red}{-1220.4} $\left(\frac{\textnormal{Exp. IH95}}{\textnormal{Exp.}}\right)$ & \textcolor{red}{-3447.4} $\left(\frac{\textnormal{Plummer R16}}{\textnormal{Exp.}}\right)$ \\ & & & & & \textcolor{red}{-1352.6} $\left(\frac{\textnormal{King IH95}}{\textnormal{Exp.}}\right)$ & \textcolor{red}{-1466.0} $\left(\frac{\textnormal{King R16}}{\textnormal{Exp.}}\right)$ \\
			\noalign{\smallskip}
			\hline     
		\end{tabular}
	\end{table*}
	
\end{document}